\documentclass[iop]{emulateapj}
\slugcomment{{\sc Accepted to ApJ:} February 16, 2014}

\begin{document}

\title{The Density and Mass of Unshocked Ejecta in Cassiopeia A through Low 
Frequency Radio Absorption}


\author{Tracey DeLaney$^1$, Namir E. Kassim$^2$, Lawrence Rudnick$^3$, \sc{and} R. A. Perley$^4$}
\affil{$^1$Physics \& Engineering Department, West Virginia Wesleyan College, 
Buckhannon, WV 26201, USA; delaney\_t@wvwc.edu\\ U.S. Naval Research Laboratory, Washington, DC  20375, USA: namir.kassim@nrl.navy.mil\\ Minnesota Institute for Astrophysics, School of Physics and Astronomy, University of Minnesota, 116 Church Street SE,\\ Minneapolis, MN 55455, USA; larry@astro.umn.edu\\ National Radio Astronomy Observatory, P. O. Box O, Socorro, NM 87801, 
USA: rperley@nrao.edu}





\begin{abstract}

Characterizing the ejecta in young supernova remnants is a requisite step 
towards a better understanding of stellar evolution. In Cassiopeia~A 
the density and total mass remaining in the unshocked ejecta are 
important parameters for modeling its explosion and subsequent evolution. Low 
frequency ($<$100~MHz) radio observations of sufficient angular resolution 
offer a unique probe of unshocked ejecta revealed via free-free absorption 
against the synchrotron emitting shell. We have used the Very Large Array plus 
Pie Town Link extension to probe this cool, ionized absorber at 
9$\arcsec$ and 18\farcs5 resolution at 74~MHz. Together with higher frequency 
data we estimate an electron density of 4.2~cm$^{-3}$ and a total mass of 
0.39~M$_{\sun}$ with uncertainties of a factor of $\sim$2. This is a 
significant improvement over the 100~cm$^{-3}$ upper limit offered by infrared 
[\ion{S}{3}] line ratios from the \emph{Spitzer Space Telescope}.  Our 
estimates are sensitive to a number of factors including temperature and 
geometry. However using reasonable values for each, our unshocked mass 
estimate agrees with predictions from dynamical models. We also consider the 
presence, or absence, of cold iron- and carbon-rich ejecta and how these 
affect our calculations.  Finally we reconcile the intrinsic absorption from 
unshocked ejecta with the turnover in Cas~A's integrated spectrum documented 
decades ago at much lower frequencies. These and other recent observations 
below 100~MHz confirm that spatially resolved thermal absorption, when 
extended to lower frequencies and higher resolution, will offer a powerful 
new tool for low frequency astrophysics.

\end{abstract}



\section{Introduction}

\subsection{General Background on Cas~A}

Cassiopeia~A (Cas~A; 3C 461, G111.7-2.1) is the 2nd-youngest-known supernova 
remnant (SNR) and, at a distance of 3.4 kpc \citep{rhf95}, it lies just 
beyond the Perseus Arm of the Galaxy.  With the discovery of light echoes 
from the explosion, we now know that Cas~A resulted from a type IIb explosion 
\citep{kbu08}.  Cas~A is one of the strongest synchrotron radio-emitting 
objects in the sky and has been observed extensively with the Very Large 
Array (VLA) since its commissioning in 1980.  

The morphology of Cas~A is quite complex with structure distributed over a 
variety of spatial scales.  The terminology used to describe some of these 
structures was coined in some of the earliest papers describing the optical 
and resolved radio images \citep{vd70,ren65}.  The most prominent 
feature is the almost circular ``Bright Ring'' at a radius of 
$\approx100\arcsec$ which is generally regarded as marking the location of 
ejecta that have interacted with the reverse shock \citep[see e.g.][]{mfc04}.  
A fainter ``plateau'' of radio emission is seen out to a radius of 
$\approx150\arcsec$.  To the northeast, where the shell becomes broken, is 
the ``jet'' and extending in the opposite direction to the southwest is the 
counter-jet.  The jet and counter-jet do not represent outflow in the 
classical sense.  Instead they describe locations where the fastest-moving 
ejecta are observed well beyond the plateau and farthest from the explosion 
center \citep[see e.g.][]{hf08}.  To the southeast, iron-rich ejecta 
extend beyond the Bright Ring and into the plateau.  The jets and extended 
iron-rich structure are likely the result of an asymmetric explosion of the 
progenitor \citep[see e.g.][]{hrb00,hl12}.  The light echo data also indicate 
an asymmetric explosion \citep{rfs11}.  

\subsection{Unshocked Ejecta}

In addition to the shocked ejecta described above, there is a class of ejecta 
still interior to the reverse shock in Cas~A.  These ``unshocked ejecta'' were 
discovered via absorption of low frequency ($<$100~MHz) radio emission 
\citep{kpd95} and are also seen to radiate in the infrared in the emission 
lines of [\ion{O}{4}], [\ion{S}{3}], [\ion{S}{4}], and [\ion{Si}{2}] 
\citep{err06}.  The term ``unshocked'' is somewhat of a misnomer because all 
of the ejecta were originally shocked by the passage of the blast wave through 
the star.  However, the ejecta cooled during the subsequent expansion of the 
SNR.  What we consider to be shocked ejecta today are those ejecta that have 
crossed through the reverse shock with the term ``unshocked ejecta'' referring 
to those ejecta that are still interior to the reverse shock.  Thus the simple 
cartoon of Cas~A's structure is that of cold ejecta in the interior of a 
roughly spherical shell composed of shocked gas that radiates strongly in 
multiple bands.  At low frequencies, the radio emission from the far side of 
the shocked shell is absorbed by the cold, unshocked ejecta in the interior.  
In \S\ref{sec:geometry} we provide a more thorough description of the geometry 
assumed for our analysis.

The infrared emission from the unshocked ejecta in Cas~A occurs because they 
have been photoionized.  An analysis of the unshocked ejecta observed in the 
supernova remnant SN1006, based on calculations of photoionization 
cross-sections and Bethe parameters, showed that the unshocked ejecta are 
photoionized by two primary sources \citep{hf88}.  Ambient ultraviolet 
starlight is all that is necessary to photoionize \ion{Si}{1} and \ion{Fe}{1} 
since their ionization potentials are below the Lyman limit.  For ions with 
ionization potentials above the Lyman limit, the ultraviolet and soft X-ray 
radiation field of the shocked ejecta in SN1006 is such that species up to 
\ion{O}{3} (54.9~eV), \ion{Si}{4} (45.1~eV), and \ion{Fe}{4} (54.8~eV) can be 
photoionized.  A similar analysis was performed for Cas~A assuming 
photoionization equilibrium, abundances appropriate for a core-collapse SNR, a 
thermal bremsstrahlung spectrum for the shocked ejecta, and utilizing a simple 
one-dimensional hydrodynamic model to follow Cas~A's evolution \citep{e09}. 
This simplified model predicts that [\ion{Si}{2}] (34.8\micron) and 
[\ion{O}{4}] (25.9\micron) should be the dominant infrared lines in the 
unshocked ejecta, directly in line with observations \citep{err06}.  In 
addition, \citet{e09} predicts strong [\ion{O}{3}] (88.4\micron), which we 
will show in \S\ref{sec:composition} is present in spectra from 
the \emph{Infrared Space Observatory} \citep[\emph{ISO},][]{upc97,ds10}.

Cas~A was spectrally mapped with the \emph{Spitzer} Space Telescope and 
Doppler shifts were measured which allowed a 3D mapping of the ejecta 
distribution, including the unshocked component \citep{drs10,ird10}.  We now 
know based on the \emph{Spitzer} data that the ejecta are organized into a 
``thick disk'' structure, tilted at $\sim70\degr$ from the line-of-sight, 
providing further evidence that the explosion, or subsequent evolution of the 
SNR prior to the reverse shock encounter, must have been asymmetric.  An upper 
limit of 100~cm$^{-1}$ was determined for the electron density of the 
unshocked ejecta based on infrared [\ion{S}{3}] line ratios, but the actual 
density is likely much lower \citep{e09,srd09}.

The absorption seen in the low frequency radio observations provides a means 
to probe the density and mass of the unshocked ejecta because the free-free 
optical depth ($\tau_{\nu}$) is related to emission measure and thus density.  
\citet{kpd95} attempted to determine the total mass of the unshocked ejecta, 
but due to using a $\tau_{\nu}$ appropriate for a hydrogenic gas and a 
temperature that was too high, arrived at 19M$_{\sun}$, which is unreasonably 
large considering that the total ejecta mass is likely only 2-4~M$_{\sun}$ 
\citep{hl12}.  Given the role that the ejecta play in the evolution of SNRs, 
it is important to provide an accurate census of the total mass present, thus 
prompting a new look at the low frequency absorption analysis of \citet{kpd95}.

\subsection{Low Frequencies on the VLA}

\subsubsection{Low Frequencies on the Legacy VLA}

The upgraded Karl G. Jansky Very Large Array (VLA) primarily accesses 
frequencies above 1~GHz through its broadband Cassegrain focus systems. Its 
predecessor, hereafter the ``legacy'' VLA, also accessed two relatively narrow 
bands below 1~GHz through its primary focus systems \citep{kpe93,kle07}. These 
included the ``P~band'' and ``4~band'' systems operating at 330~MHz 
(1990-2009) and 74~MHz (1998-2009), respectively. Both systems provided 
sub-arcminute resolution imaging and were widely used over their lifetimes.  
These systems were removed during the VLA upgrade and have been only
recently replaced with a new ``Low Band'' receiving system \citep{ckh11}. The 
first call for proposals using the 330 MHz band of this new system was issued 
by NRAO in February, 2013.

\subsubsection{Pie Town Link}

The legacy 74~MHz and 330~MHz VLA systems achieved their maximum angular 
resolution of $\sim20\arcsec$ and $\sim6\arcsec$ in the A~configuration 
(maximum baseline $\sim$36~km), respectively. An 8-antenna prototype of the 
74~MHz system was used to observe Cas~A \citep{kpd95} early on, adding to the 
body of VLA work extending from 330~MHz to higher frequencies.  As the 
resolution was still relatively poor compared to shorter wavelengths, NRL 
and NRAO added a 74~MHz feed system to the Pie Town antenna\footnote{The Pie 
Town antenna already had a permanent 330~MHz feed as part of the VLBA.}, 
utilizing an optical fiber link connecting the innermost Very Large Baseline 
Array (VLBA) antenna to the legacy VLA\footnote{The optical fiber link was 
experimental and is no longer active.}. With a maximum baseline of 73~km, this 
improved the angular resolution at 74~MHz and 330~MHz by a factor of two to 
approximately 9$\arcsec$ and 3$\arcsec$ respectively.  Cas~A was thereafter 
reobserved with the full (27-antenna) 74~MHz and the 330~MHz legacy systems 
using the Pie Town link capability in August 2003.

\subsection{This Paper}

In this paper, we report on legacy VLA observations in 1997-1998 at 
frequencies of 5~GHz, 1.4~GHz, 330~MHz, and 74~MHz, and on the Pie Town link, 
observations at frequencies of 330~MHz and 74~MHz in 2003.  We use our derived  
images to determine the degree of free-free absorption present, applying the 
appropriate formula for when the composition of the gas is not dominated by 
hydrogen.  Finally, we calculate the density and mass of the unshocked ejecta 
and discuss the implications of our results. 

\section{Observations and Imaging}

\subsection{1997-1998 and 2003 Legacy VLA Observations and Calibration}

Observations were made with the legacy VLA in 1997 and 1998 using all four 
configurations from the most extended A-configuration with a maximum antenna 
separation of 36.4~km to the most compact D-configuration with a shortest 
separation of 35~m.  These are summarized in Table~\ref{e1table}.  Data were 
taken at 4410.0, 4640.0, 4985.0, and 5085.0~MHz (hereafter called 5~GHz or 
C~band), at 1285.0, 1365.1, 1464.9, and 1665.0~MHz (hereafter called 1.4~GHz 
or L~band), at 327.5 and 333.0~MHz (hereafter called 330~MHz or P~band), and 
at 73.8~MHz (hereafter called 74~MHz or 4~band).  The 5~GHz and 1.4~GHz data 
were taken in continuum mode with pairs of frequencies observed 
simultaneously.  The purpose of observing at multiple 5~GHz and 1.4~GHz bands 
is to improve sensitivity while avoiding bandwidth smearing.  The 330~MHz and 
74~MHz data were taken in spectral line mode. Note that the D-configuration 
data taken at 330~MHz and 74~MHz had insufficient resolution to be useful for 
this paper and were not utilized.

In order to increase the spatial resolution of the lowest frequency images, 
further observations at 74~MHz and 330~MHz were taken with the legacy VLA in 
A-configuration and utilizing the Pie Town (PT) link (hereafter called A+PT).  
We were allocated an 18-hour observing block in Aug 2003 with 12.2 hours on 
Cas~A and the remainder of the time on calibrators and other sources of 
interest at low frequencies.  The 2003 observations are also summarized in 
Table~\ref{e1table}.

\begin{deluxetable}{lcccc} 
\tablecaption{\label{e1table} Cas~A 1997-1998 and 2003 Legacy VLA Observations\tablenotemark{a}}
\tabletypesize{\small}
\tablewidth{0.46\textwidth}
\tablehead{
\colhead{} & 
\colhead{} & 
\colhead{} & 
\colhead{} & 
\colhead{\hspace*{-1mm}On-Source Time\hspace*{-1mm}} 
\\
\colhead{} & 
\colhead{} & 
\colhead{} & 
\colhead{\hspace*{-2mm}Bandwidth\tablenotemark{c}\hspace*{-2mm}} & 
\colhead{per Band}
\\
\colhead{Date} & 
\colhead{\hspace*{-2mm}Configuration\hspace*{-2mm}} & 
\colhead{Band\tablenotemark{b}} & 
\colhead{(MHz)} & 
\colhead{(minutes)}
}
\startdata
1997 May & B & 4 & 1.5625 & 33 \\
 & & P & 3.125 & 67 \\
 & & L & 12.5 & 241 \\
 & & C & 12.5 & 293 \\
1997 Sep & C & 4 & 1.5625 & 24 \\
 & & P & 3.125 & 24 \\ 
 & & L & 12.5 & 144 \\ 
 & & C & 12.5 & 142 \\ 
\hspace*{-2mm}1997 Nov, Dec\hspace*{-2mm} & D & 4 & 1.5625 & 14 \\
 & & P & 3.125 & 14 \\
 & & L & 12.5 & 84 \\
 & & C & 12.5 & 123 \\
1998 Mar & A & 4 & 1.5625 & 61 \\
 & & P & 3.125 & 61 \\
 & & L & 6.25 & 464 \\
 & & C & 6.25 & 467 \\
2003 Aug & A+PT & 4 & 1.5625 & 732 \\
 & & P & 3.125 & 732 \\
\enddata
\tablenotetext{a}{Original proposal codes were AR378 (1997-1998) and AD480 
(2003).}
\tablenotetext{b}{These are telescope time on source at each band.  For C and 
L~band observations, this is the combined time for both frequency pairs since 
pairs of frequencies are observed simultaneously.}
\tablenotetext{c}{These bandwidths are for each frequency observed.}
\end{deluxetable}

Standard calibration procedures at 1.4 and 5~GHz were used for the data as 
described in the AIPS 
Cookbook\footnote{\url{http://www.aips.nrao.edu/cook.html}}.  Primary flux 
calibration was based on the source 3C~48, the polarization calibrator was 
3C~138, and the phase calibrator was J2355+4950.  After initial calibration, 
multiple passes of self-calibration and imaging were performed on the Cas~A 
data to improve the antenna phase and gain solutions.

The observations at 330~MHz and 74~MHz were performed in spectral line mode 
due to the presence of narrow-band radio frequency interference (RFI).  Images 
of Cygnus~A at 330~MHz and 74~MHz, after normalization to the \citet{bgp77} 
absolute flux density scale, were used for both flux and bandpass 
calibration.  RFI was flagged out and the data were combined into a single 
channel at each band.  Iterative cycling between self calibration and imaging 
was then used to improve the phase and gain solutions for Cas~A and to derive 
the final images. 

74~MHz legacy VLA data, especially for the B, A, and A+PT link configurations, 
require corrections for ionospheric phase variations that can be rapid and 
severe over baselines longer than several kilometers \citep{kle07}. 
Fortunately, Cas~A so completely dominates the visibility phase measurements 
that straightforward self-calibration, yielding one phase correction per 
antenna, is perfectly capable of measuring the ionospheric effects. Moreover, 
the signal-to-noise is sufficiently high that the fluctuations can be tracked 
and corrected at the native sampling rate of 6.7 seconds.

Standard 74~MHz and 330~MHz data reduction also typically requires 
wide-field, multi-faceted imaging of spectral line data, to address 
non-coplanar baseline imaging and bandwidth smearing, respectively 
\citep{kle07}.  Fortunately, Cas~A is of small angular size (only 
$\sim300\arcsec$ across) such that there is no need for implementing 
angular-dependent gain solutions nor dealing with the non-coplanar array.  
Furthermore, Cas~A is so strong that the contribution of other sources in the 
primary beams at both frequencies is negligible, and hence neither technique 
was required, greatly simplifying the data reduction.

\subsection{Total Intensity Images}
\label{lfims}

\begin{deluxetable*}{cccrrrrrr} 
\tablecaption{\label{fluxes} Final Image Statistics}
\tablecolumns{5}
\tabletypesize{\small}
\tablewidth{1\textwidth}
\tablehead{
\colhead{} & 
\colhead{} & 
\colhead{$u$-$v$ range} & 
\colhead{$\theta_{LAS}$\tablenotemark{a}} & 
\colhead{FOV\tablenotemark{b}} & 
\colhead{Noise} &
\colhead{Dynamic\tablenotemark{c}} & 
\colhead{$S_{\nu}$\tablenotemark{d}} &
\colhead{$\delta S_{\nu}$\tablenotemark{e}} 
\\
\colhead{Band} & 
\colhead{Epoch} & 
\colhead{(k$\lambda$)} & 
\colhead{(arcsec)} & 
\colhead{(arcmin)} & 
\colhead{(mJy bm$^{-1}$)} & 
\colhead{Range} &
\colhead{(Jy)} &
\colhead{(Jy)}
}
\startdata
\multicolumn{9}{c}{2\farcs5 Resolution} \\ \tableline
P & 2003 & 0.7-81 & 170 & 150 & 5.5 & 40 & 6217 & 124 \\
L & 1997/8 & 0.5-81 & 300 & 30 & 1.8 & 193 & 2204 & 44 \\
C & 1997/8 & 0.5-81 & 300 & 7.5 & 3.2 & 194 & 809 & 16 \\ \tableline
\multicolumn{9}{c}{9\arcsec~Resolution} \\ \tableline
4 & 2003 & 0.7-18 & 170 & 600 & 190 & 368 & 18951 & 379 \\
P & 2003 & 0.7-18 & 170 & 150 & 26 & 135 & 6217 & 124 \\ \tableline
\multicolumn{9}{c}{18\farcs5 Resolution} \\ \tableline
4 & 1997/8 & 0.12-9 & 800 & 600 & 387 & 513 & 18555 & 371 \\
P & 1997/8 & 0.5-9 & 300 & 150 & 94 & 267 & 6185 & 124 \\
 & & (0.12-9) & & & & & (5859)\tablenotemark{f} \\
L & 1997/8 & 0.12-9 & 800 & 30 & 21 & 249 & 2179 & 44 \\
\enddata
\tablenotetext{a}{The largest angular scale visible to the array.  Note that 
Cas~A is 300$\arcsec$ (5$\arcmin$) in diameter.}
\tablenotetext{b}{The field of view (FOV), or primary beam, of the array set 
by the single dish aperture and the observing band.  At 5~GHz, the FOV is only 
1.5 times the size of Cas~A.}
\tablenotetext{c}{dynamic range=(image maximum)/(3-$\sigma$)}
\tablenotetext{d}{Integrated flux densities are reported at the frequencies of 
4.64~GHz, 1.285~GHz, 330~MHz, and 73.8~MHz.}
\tablenotetext{e}{Based on a 2\% uncertainty in the absolute flux 
calibration \citep{pt91}.}
\tablenotetext{f}{Integrated flux density for 330~MHz image using 
0.12~k$\lambda$ as the minimum $u$-$v$ distance to show the magnitude of the 
Van Vleck bias.}
\end{deluxetable*}

\begin{figure}[b]
\epsscale{1.2}
\plotone{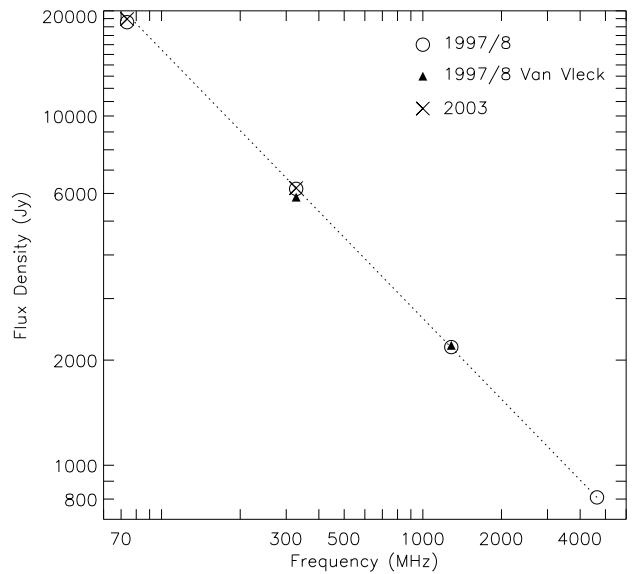}
\vspace{-2mm}
\caption{Integrated flux density measurements at each of the four observing 
bands and using the $u$-$v$ ranges indicated in Table~\ref{fluxes}.  The error 
in flux density is 2\%, which is much smaller than the plotting symbols.
The filled triangles indicate the flux densities of the 1997-1998 1.4~GHz and 
330~MHz images using their native minimum $u$-$v$ distances to demonstrate the 
magnitude of the Van Vleck bias in each.  The dotted line represents a 
spectral index of -0.77.
\label{vvcorr}}
\end{figure}

We present in this section sets of ``matched'' images at 2\farcs5, 9\arcsec, 
and 18\farcs5 resolution.  The individual configuration data at each band were 
concatenated and images made with matched beam sizes.  Minimum and maximum 
spatial frequency ($u$-$v$) ranges were used to match spatial sampling with 
the exception of the 2\farcs5 and 18\farcs5 330-MHz images which have a 
different minimum $u$-$v$ cutoff than the other images due to the native 
minimum $u$-$v$ distance for the 2003 A+PT data and to avoid the Van Vleck 
bias for the 1997-1998 data, as discussed below.  The AIPS maximum-entropy 
deconvolution routine VTESS was used to restore the total intensity images.  
The default image for VTESS was a 5~GHz image of appropriate spatial 
resolution.  Using the same default image in VTESS for all of the bands 
essentially forces the images to look as much alike as possible, with any 
resulting differences being due to the requirements of the data.  Initial 
zero-spacing flux guesses were supplied to VTESS based on the 5~GHz integrated 
flux density and Cas~A's average spectral index of -0.77.  The ``negative 
flux'' option was used so that VTESS was not required to get within 5\% of the 
flux estimate.  The standard correction for primary beam attenuation was 
applied to all of the final images.  Since there were four images at 1.4~GHz 
and four images at 5~GHz, the images for each band needed to be normalized in 
flux density and averaged together.  While we could have concatenated the 
final calibrated $u$-$v$ datasets and then performed the imaging, we chose to 
normalize the images directly.  There are two primary reasons for this choice.  
First, the slightly different primary beam correction at each band introduces 
a small, but not insignificant error.  Second, it is generally not a good idea 
to force the $u$-$v$ data to the same flux density, and grid on one plane for 
objects, such as Cas~A, that have significant spectral gradients because each 
$u$-$v$ cell has a slightly different visibility amplitude at each band.  
Furthermore, since the individual 5~GHz and 1.4~GHz bands are close together, 
the relative calibration uncertainty is large enough that Cas~A's well-known 
average spectral index cannot be used to scale between them.  This 
necessitates doing the concatenation and normalization in the imaging plane.  
We determined the average scaling factor between each set of 5~GHz and 1.4~GHz 
images using the radial surface brightness profiles extracted from a series of 
20\degr~wedges.  The 1.4~GHz images were scaled to the 1.285 GHz image and the 
5~GHz images were scaled to the 4.64~GHz image.  In Table~\ref{fluxes}, we 
summarize the integrated flux densities, noise figures, and $u$-$v$ ranges for 
the final images and in Figure~\ref{vvcorr} we plot the integrated radio 
spectrum of Cas~A.  The uncertainty in the integrated flux density at each 
band is calculated based on a 2\% uncertainty in the absolute flux scale of 
the VLA \citep{pt91}.  

At 330~MHz, legacy VLA observations of Cas~A suffer from the Van Vleck bias 
\citep{vm66} which depends very strongly and non-linearly on the correlated 
flux density and also affects phase since the degradation operates 
independently on the real and imaginary portions of the visibility separately.
Thus, short antenna spacings report a reduced flux density which skews the 
final reconstructed flux density of the image by about 6\% as shown in 
Table~\ref{fluxes} and Figure~\ref{vvcorr}.  An effect this large would create 
a significant bias in spectral index and absorption measurements.  For the 
A+PT 330~MHz data taken in 2003, the minimum $u$-$v$ distance is 
0.7~k$\lambda$, which is long enough to avoid the Van Vleck bias.  In order to 
avoid this bias for the 1997-1998 330~MHz image, which has data from shorter 
baselines, we impose a 0.5~k$\lambda$ minimum to the $u$-$v$ range.  This 
limit happens to correspond to the minimum baseline for the 5~GHz data in 
D-configuration and results in a completely negligible Van Vleck bias 
\citep{tms07}.  Unfortunately, variations in spatial sampling between images 
can affect spectral index measurements.  Different deconvolution 
methods provide different abilities to ``fill in'' the holes in the $u$-$v$ 
sampling.  Strict overlap or matching of $u$-$v$ coverage is not required to 
get a good image, especially for an exceptionally bright source such as 
Cas~A.  Furthermore, good deconvolution algorithms are able to recover the 
short-spacing flux, especially when they permit a total flux estimate and 
allow a default image, as VTESS does.  

The Van Vleck bias also should affect the 1.4~GHz images based on the 
correlation coefficient \citep{tms07}, however we note only a 1\% difference 
in the total flux density when using 0.5~k$\lambda$ vs. 0.12~k$\lambda$ as the 
minimum of the $u$-$v$ range (0.12~k$\lambda$ corresponds to the minimum 
baseline for the 1.4~GHz data in D-configuration.)  The 1\% flux density 
difference at 1.4~GHz is small enough that other factors besides the Van Vleck 
bias might be accountable.  

The integrated spectrum of Cas~A plotted in Figure~\ref{vvcorr} shows that the 
spectral index ($\alpha_{\nu_1}^{\nu_2}\equiv \log(S_1/S_2)/\log(\nu_1/\nu_2)$) is 
constant from 5~GHz up to 330~MHz ($\alpha_{330}^{1.4}=-0.76\pm0.01$ 
and $\alpha_{1.4}^{5}=-0.77\pm0.01$) when using a minimum $u$-$v$ cutoff for the 
330~MHz image.  From 330~MHz to 74~MHz, the integrated spectrum flattens 
slightly ($\alpha_{74}^{330}=-0.74\pm0.01$).  The uncertainties in spectral 
index are calculated based on a formal propagation of the uncertainty in 
integrated flux density, and thus the spectral flattening at 74~MHz is 
statistically significant.  The spectral index between 5~GHz and 330~MHz is 
the same as that reported by \citet{bgp77} for epoch 1980 but the spectral 
index between 330~MHz and 74~MHz is steeper than the 1980 result.  
\citet{bgp77} derived a frequency-dependent secular decrease in flux density 
which would flatten the spectral index over time.  Newer low frequency 
observations have shown that there is no frequency dependence to the secular 
decrease and the overall rate is about 8\% yr$^{-1}$ \citep{r90,hk09}.  The 
integrated flux densities we report in Table~\ref{fluxes} at 5~GHz, 1.4~GHz, 
and 330~MHz are only $\sim$3\% less than the 1980 values from \citet{bgp77} 
and the 74~MHz integrated flux densities are largely consistent with those 
reported by \citet{hk09}.  There is no significant change in integrated flux 
density between 1997-1998 and 2003.  Given the differences in aperture 
coverage, constraints imposed during image reconstruction, and flux standards 
applied for the different data sets, as well as the short-term variability 
observed in Cas~A \citep{ar95,hk09}, our flux densities are in line with 
expectations.  Thus we argue that there is no significant curvature in Cas~A's 
integrated spectrum between 5~GHz and 330~MHz but there is a significant 
flattening of the integrated spectrum at lower frequencies. 

We did not concatenate the 1997-1998 and 2003 data sets in order to make 
images containing data from both epochs because of two complicating factors. 
The first is the non-negligible proper-motion induced shift, due to expansion, 
across the epochs, and the second are the variations in the brightness of 
compact features over time.  \citet{ar95}, for example, found that two-thirds 
of the compact radio features in Cas~A have brightness changes in the range 
of -2.2\% yr$^{-1}$ to +5.2\% yr$^{-1}$ and that the compact features were on 
average brightening at the rate of 1.6\% yr$^{-1}$.  At low spatial 
resolution, the proper motions and small-scale brightness changes are 
mitigated, but then there is no reason to combine the A+PT data with the 
lower resolution data except to improve signal-to-noise, which as we will 
demonstrate below, was not necessary for the analysis at 18\farcs5 resolution.

\begin{figure}[t!]
\centering
\includegraphics[width=0.4\textwidth]{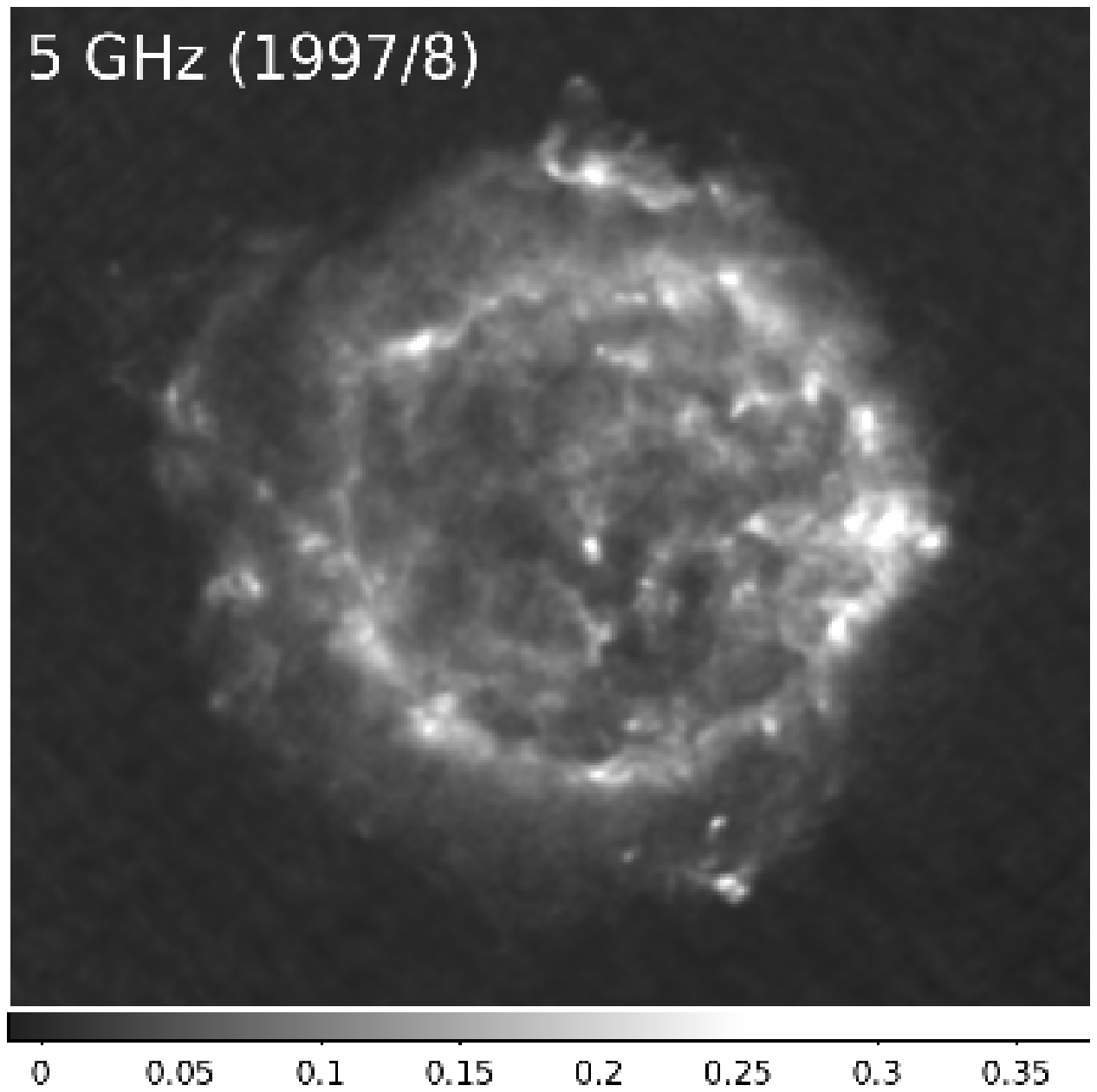}\\
\includegraphics[width=0.4\textwidth]{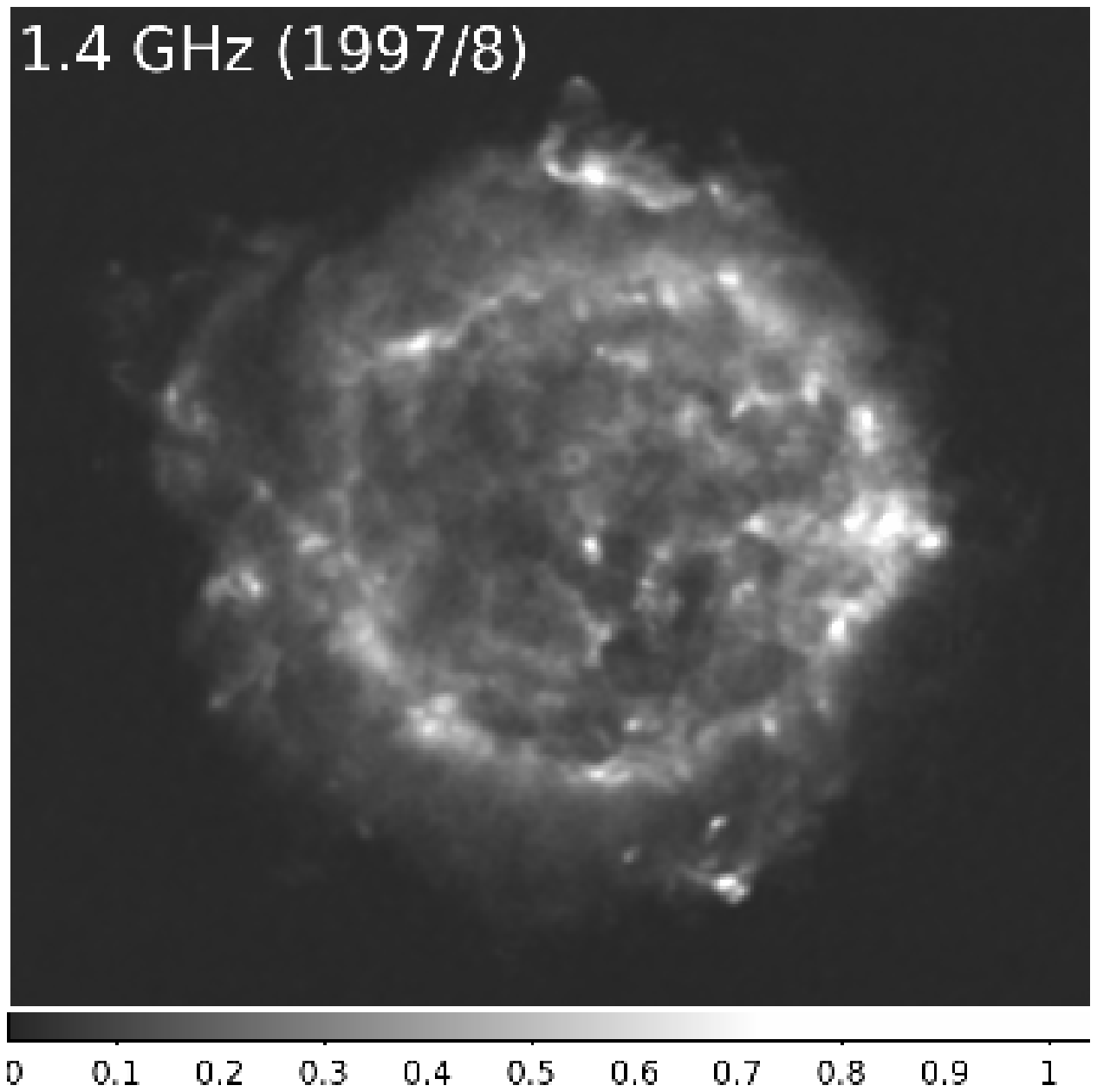}\\
\includegraphics[width=0.4\textwidth]{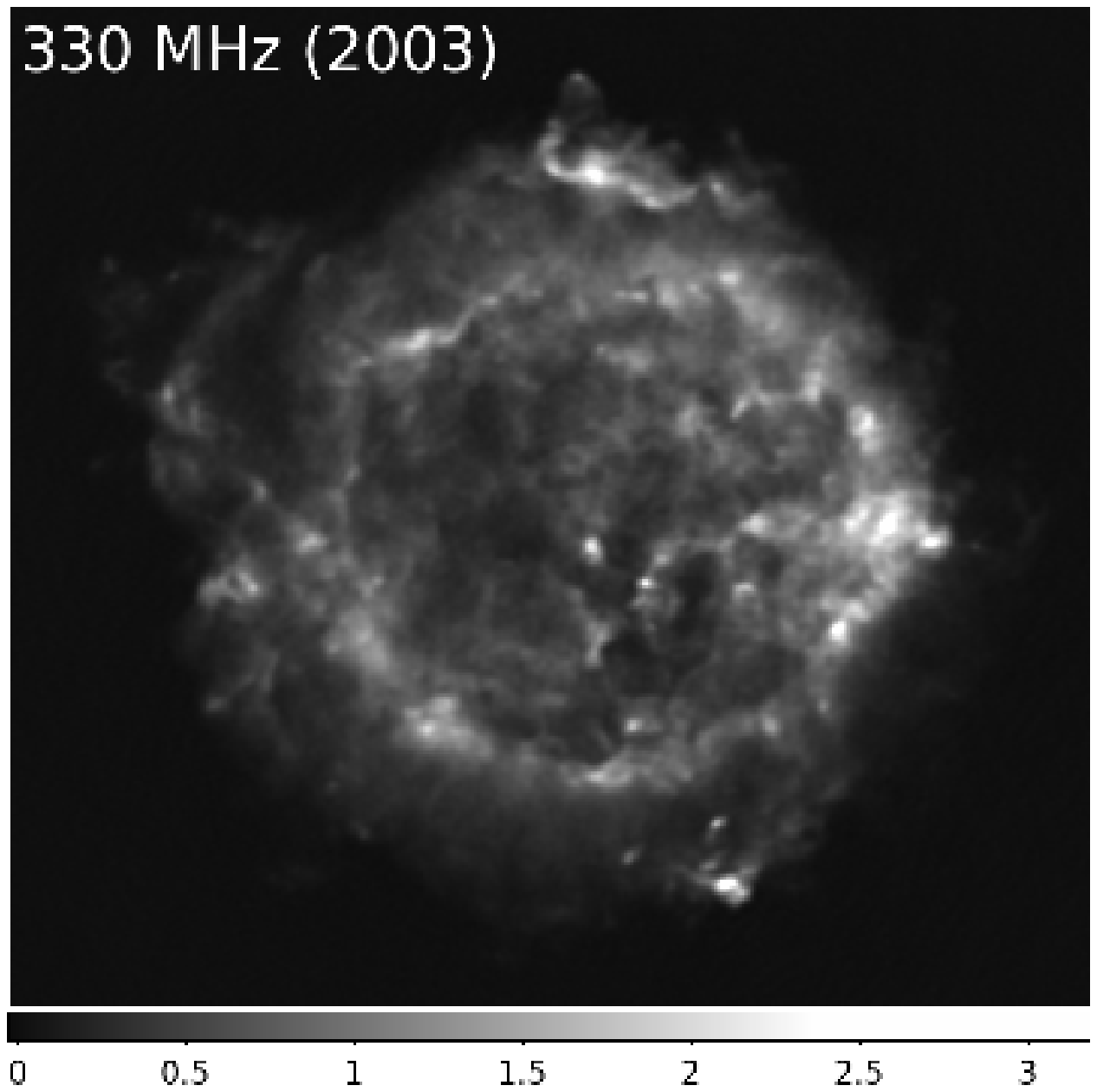}
\caption{5~GHz (top) and 1.4~GHz (middle) images of Cas~A from 1997-1998 and 
330~MHz image (bottom) from 2003.  The resolution is 2\farcs5 and the flux 
density scales are in Jy~bm$^{-1}$. The $u$-$v$ ranges used to construct these 
images and the noise and integrated flux densities are reported in 
Table~\ref{fluxes}.  
\label{2asec}}
\end{figure}

\subsubsection{2\farcs5 Resolution Images}

Figure~\ref{2asec} shows the final 1997-1998 5~GHz and 1.4~GHz images and the 
2003 330~MHz image at 2\farcs5 resolution.  With the improved resolution using 
the Pie Town link, we can see that the 330~MHz image captures the same 
large- and small-scale features as at higher frequencies, thus supporting the 
constant spectral index found from 5~GHz to 330~MHz.  The dynamic range of the 
330~MHz image is on par with the 1.4~GHz image and noticeably better than the 
5~GHz image.  The relatively poor dynamic range of the 5~GHz image is partly 
because the primary beam of the VLA at 5~GHz is only 1.5 times larger than 
Cas~A, as indicated in Table~\ref{fluxes}.  The high quality of the 330~MHz 
image demonstrates how well the AIPS task FLGIT excises RFI and the power of 
self-calibration for dealing with ionospheric effects.  

\subsubsection{9\arcsec~Resolution Images}

\begin{figure}
\centering
\includegraphics[width=0.4\textwidth]{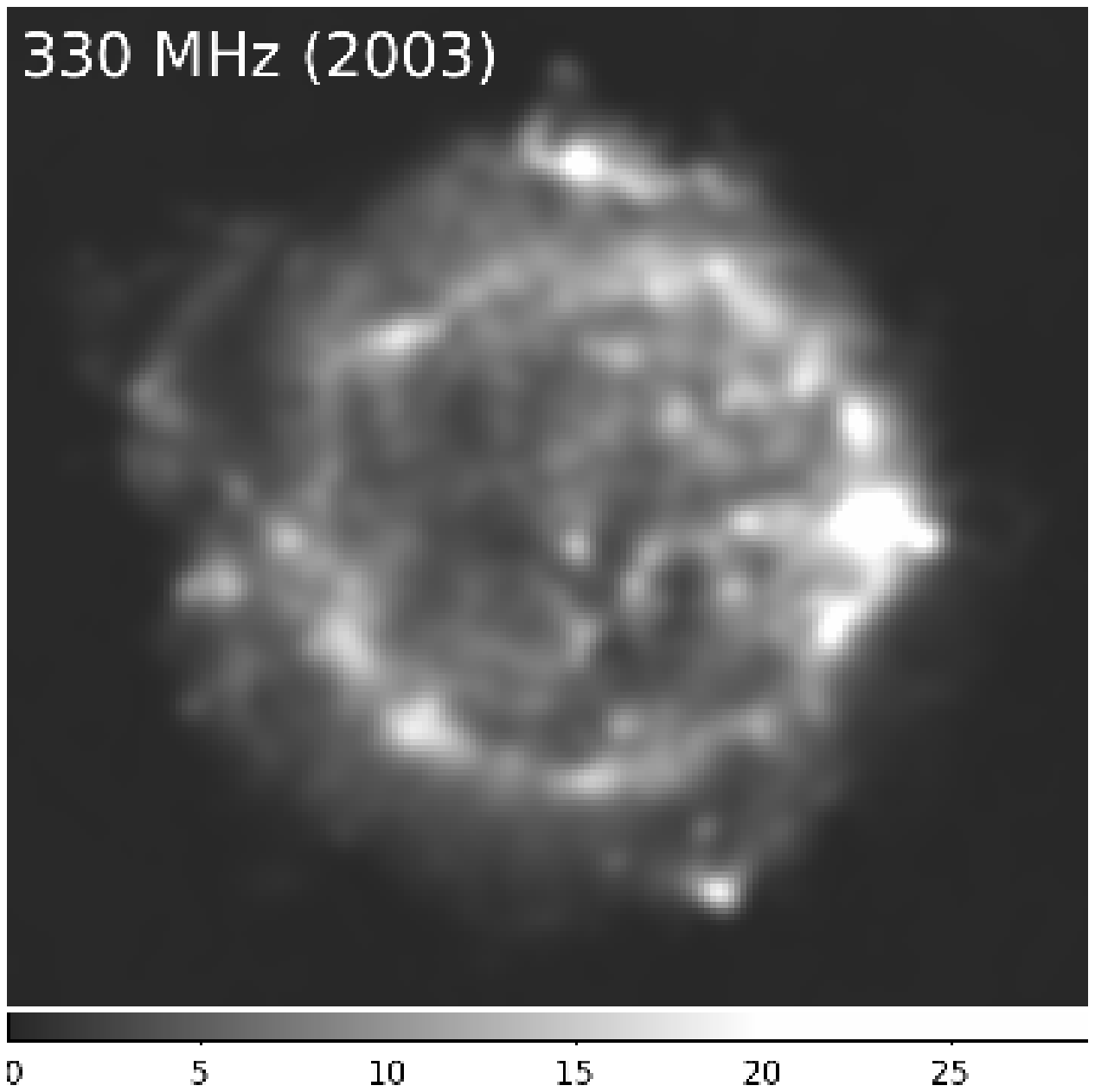}\\
\includegraphics[width=0.4\textwidth]{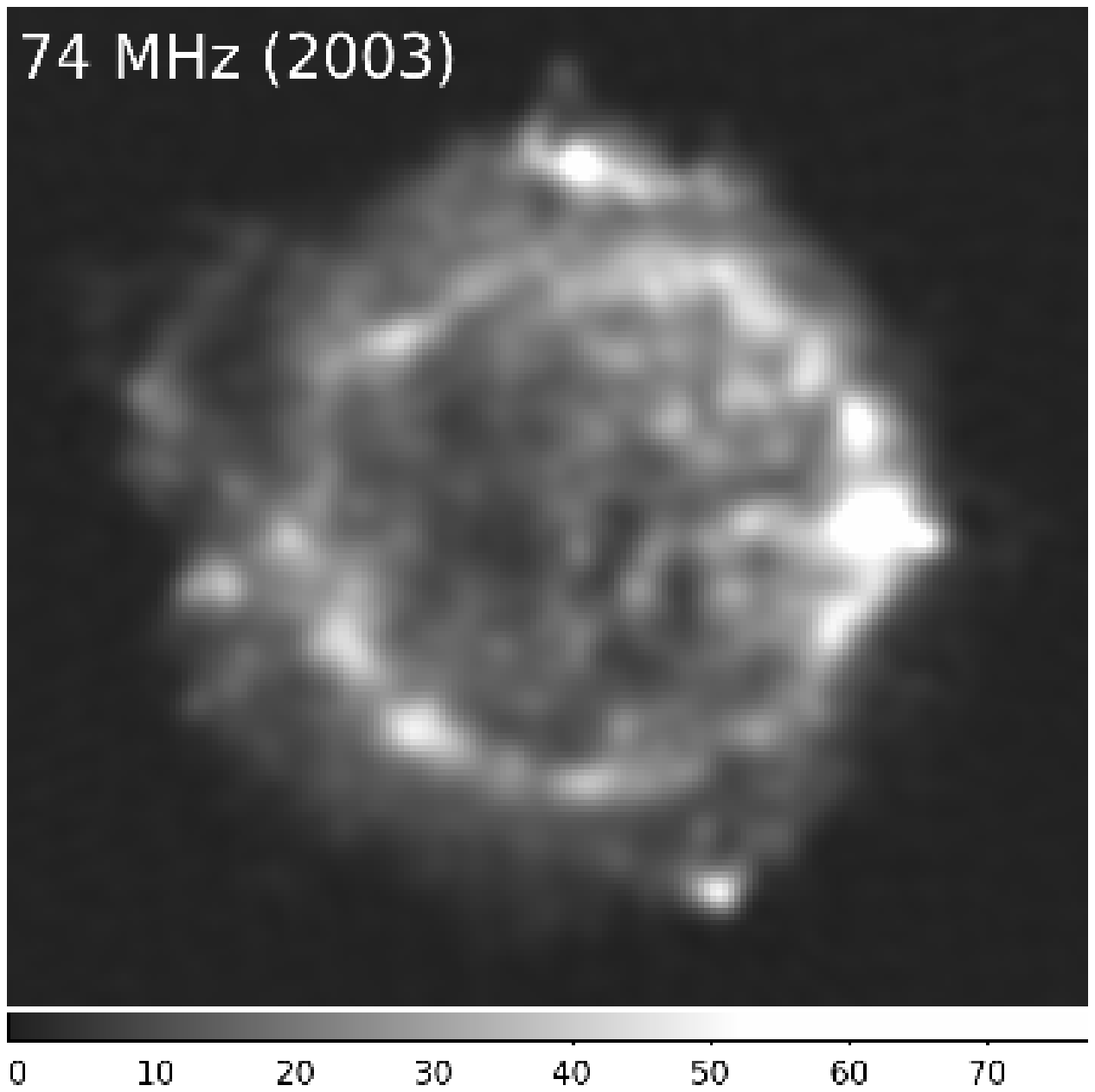}\\
\includegraphics[width=0.43\textwidth]{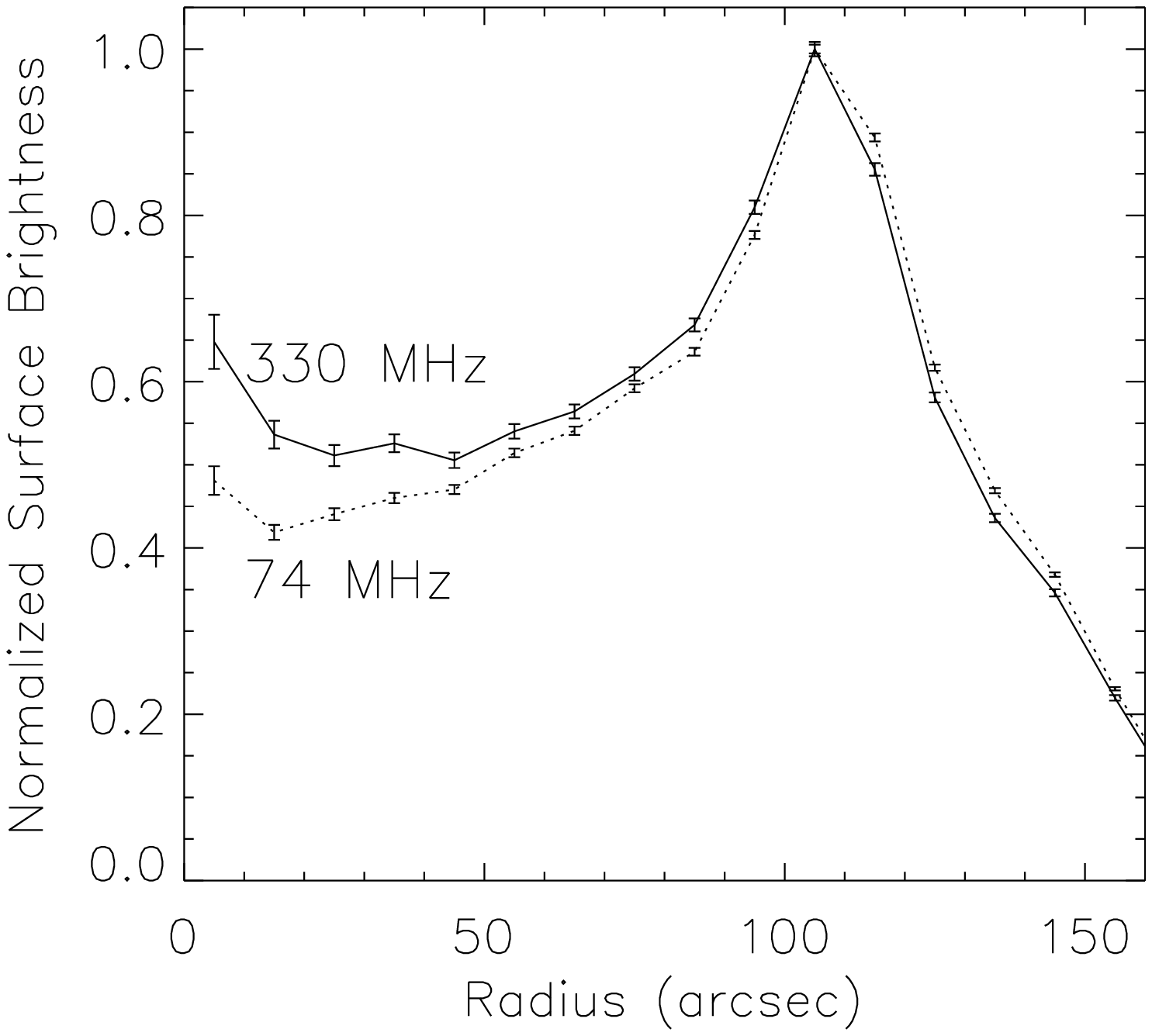}
\caption{Top and middle: 2003 A+PT~configuration images of Cas~A at 330~MHz 
and 74~MHz.  The resolution is 9\arcsec.  The flux density scales are in 
Jy~bm$^{-1}$.  The $u$-$v$ ranges used to construct these images and the noise 
and integrated flux densities are reported in Table~\ref{fluxes}.  Bottom: 
Angle-averaged radial surface brightness profiles of the 330~MHz and 74~MHz 
images normalized to their peak values. 
\label{pt9asec}}
\end{figure}

Figure~\ref{pt9asec} shows the final 2003 330~MHz and 74~MHz images at 
9\arcsec~resolution, which is the best resolution achieved with the A+PT link 
at 74~MHz.  The dynamic range on the 330~MHz image is about 3 times better 
than for the 74~MHz image.  Clumpy structure is observed on a variety of 
spatial scales down to the resolution limit.  Comparing to the 2\farcs5 
resolution images in Figure~\ref{2asec}, it is clear that the same synchrotron 
features are responsible for the emission at 74~MHz.  The bottom panel of 
Figure~\ref{pt9asec} is a plot of the angle-averaged radial surface brightness 
profiles of the 330~MHz and 74~MHz images.  The profiles have been normalized 
to their peak values.  The emission in the center of the 74~MHz image is 
noticeably fainter than expected based on the appearance of the 330~MHz image 
and therefore contributes to the flattening of the integrated spectrum from 
330~MHz to 74~MHz observed in Figure~\ref{vvcorr}.  

\subsubsection{18\farcs5 Resolution Images}

\begin{figure*}
\centering
\includegraphics[width=0.4\textwidth]{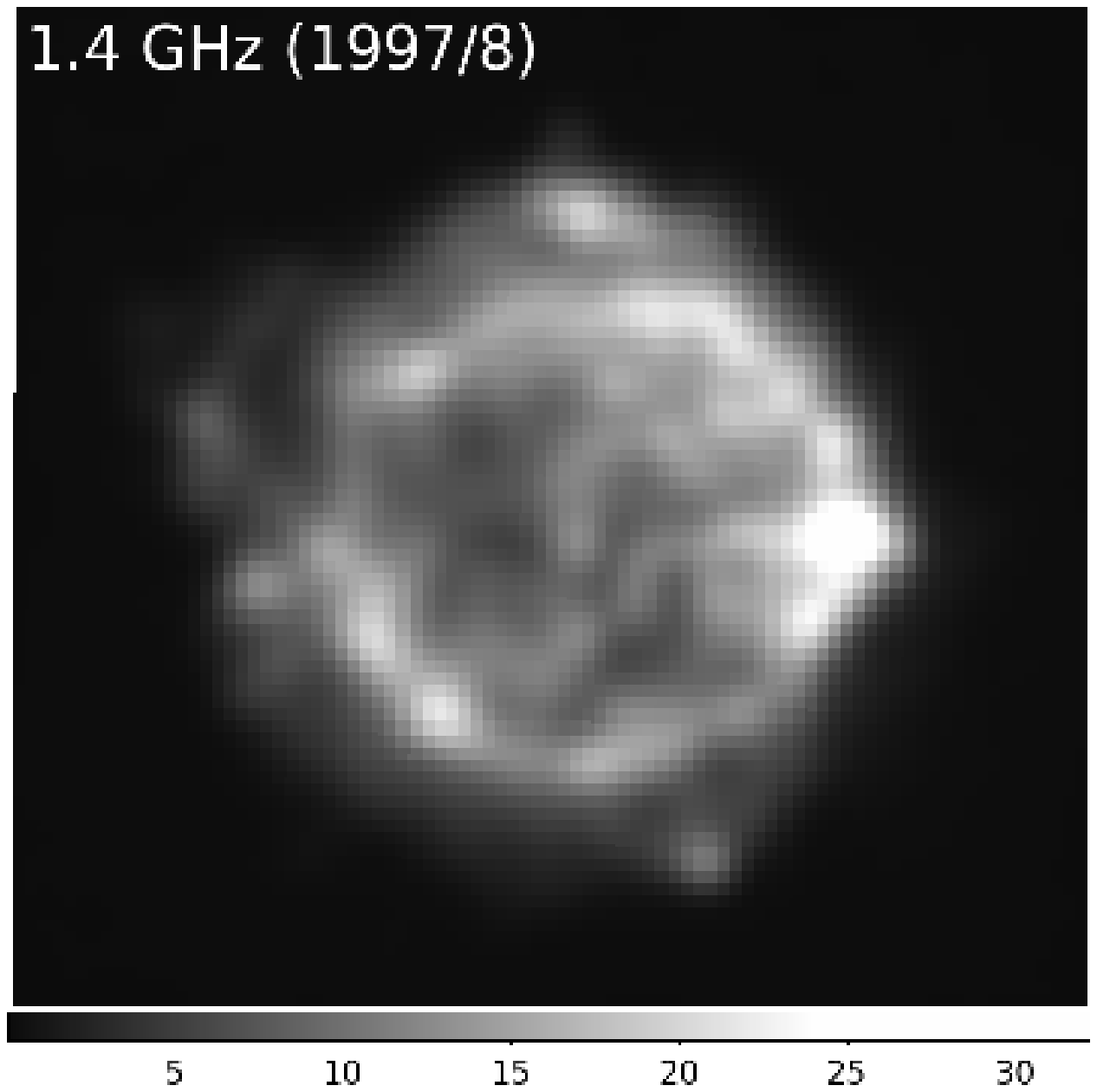}
\includegraphics[width=0.4\textwidth]{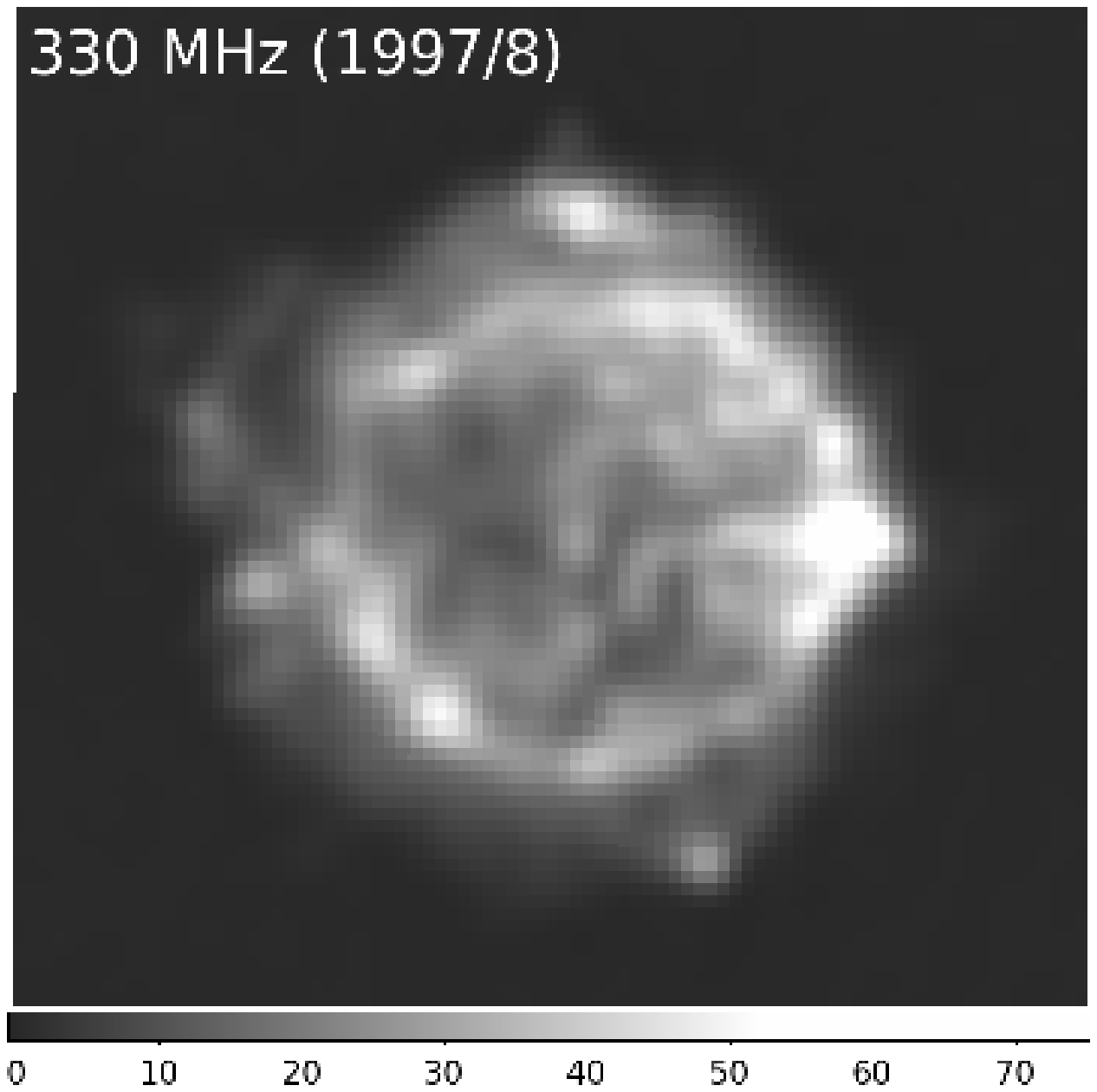}\\
\includegraphics[width=0.4\textwidth]{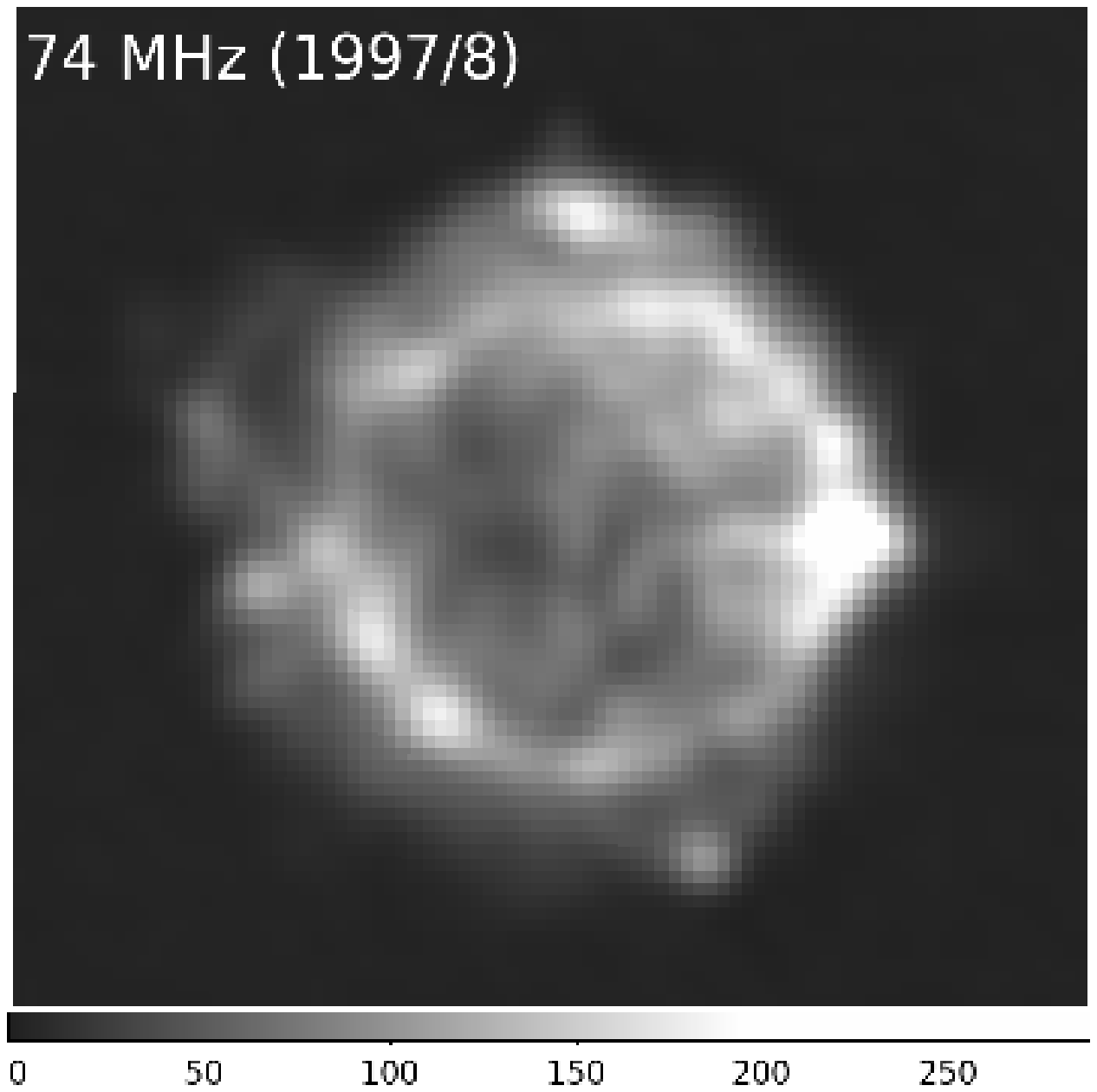}
\includegraphics[width=0.4\textwidth]{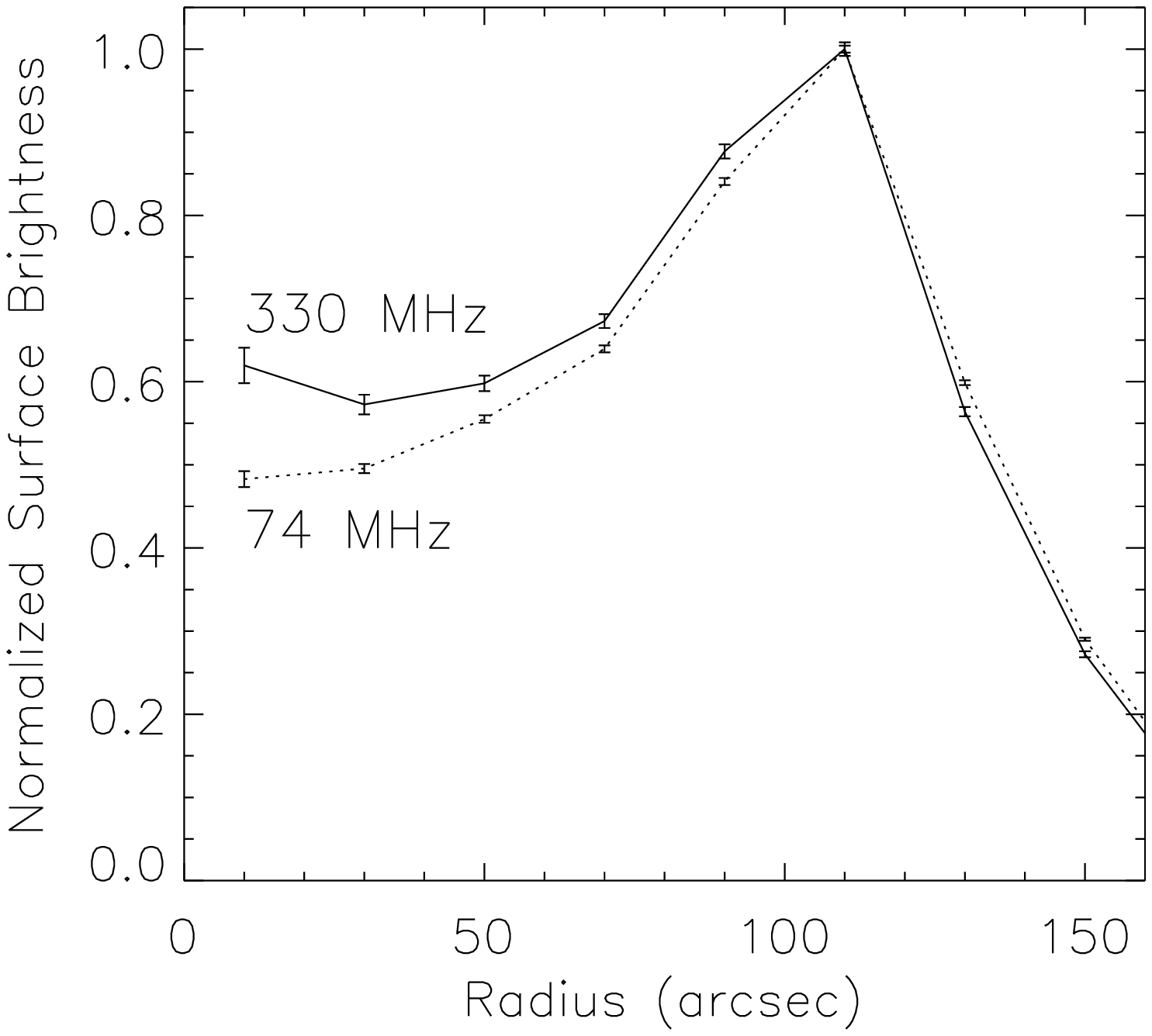}\\
\caption{1.4~GHz (top left), 330~MHz (top right), and 74~MHz (bottom left) 
images of Cas~A from 1997-1998.  The resolution is 18\farcs5 and the flux 
density scales are in Jy~bm$^{-1}$. The $u$-$v$ ranges used to construct 
these images and the noise and integrated flux densities are reported in 
Table~\ref{fluxes}.  The angle-averaged radial surface brightness profiles of 
the 330~MHz and 74~MHz normalized to their peak values are shown at bottom 
right.  \label{18asec}}
\end{figure*}

Figure~\ref{18asec} shows the final 1.4~GHz, 330~MHz and 74~MHz images at 
18\farcs5 resolution from the 1997-1998 observations.  This represents the 
best resolution attained with the legacy VLA in A-configuration at 74~MHz.  The 
dynamic range on the 1.4~GHz image is about 2 times better than for the 
330~MHz and 74~MHz images.  The 74~MHz image at this lower resolution has a 
dynamic range about twice that at higher resolution using the A+PT link.  The 
higher dynamic range allows us to better probe the large-scale diffuse 
emission from Cas~A.  The absorption of emission from the center of Cas~A is, 
again, clearly observed.

\section{Spectral Index Imaging Analysis}

\subsection{Spectral Index at 9\arcsec~Resolution}

The 2003 330~MHz and 74~MHz images at 9\arcsec~resolution were combined to 
form the spectral index image shown in Figure~\ref{ptspix}.  Only spectral 
index data from regions greater than 10-$\sigma$ on the 330~MHz image were 
retained.  The range of spectral indices at 9\arcsec~resolution is 
$\alpha_{74}^{330}\approx$-0.95 to -0.35, with the flattest spectrum 
corresponding to the bright clump in the center of Cas~A.  \citet{arl91} 
report an average $\alpha_{1.4}^{5}\approx$-0.78 for the collection of clumps 
in that location, thus a significant spectral flattening has occurred.  

The abundance of clumps with flat (blue) spectral indices in the center are 
not necessarily an indication of absorption.  Many of the central clumps are 
known to have flat spectral indices based on images at higher frequencies 
\citep{arl91}.  In order to assess the degree of absorption in Cas~A, we need 
to look at the difference in spectral index from high frequencies to low 
frequencies.  Since we did not take higher frequency data with the 2003 
observations, and the proper motion and small-scale brightness variations 
across epochs would be evident at 9\arcsec~resolution, we must work with our 
lower resolution images.

\subsection{Spectral Index at 18\farcs5 Resolution}

\begin{figure}[t!]
\epsscale{1.17}
\plotone{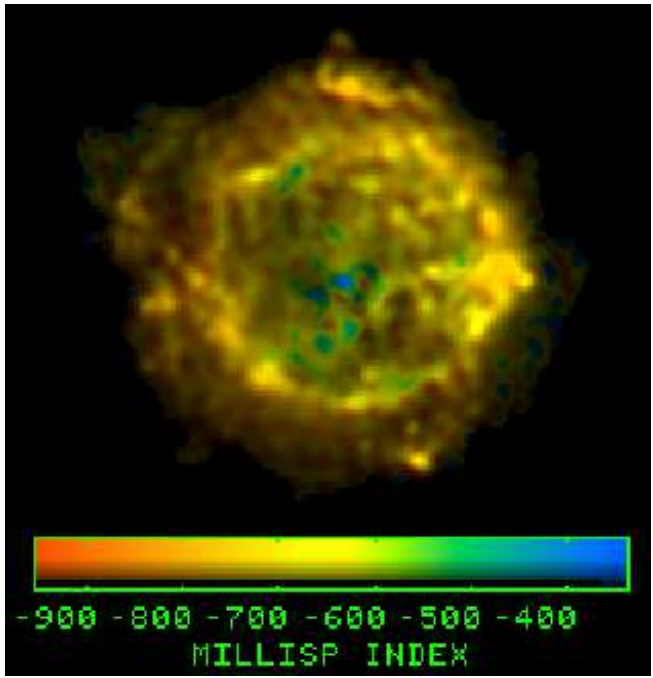}
\vspace{-1mm}
\caption{Spectral index image between 74~MHz and 330~MHz made from the 2003 
data set.  Only spectral index data from regions greater than 10-$\sigma$ on 
the 330~MHz image are shown.  The 330~MHz image is used as a brightness 
channel to illuminate the spectral index colors.  The resolution is 9\arcsec. 
\label{ptspix}}
\end{figure}

\begin{figure}[t!]
\centering
\includegraphics[height=0.4\textwidth]{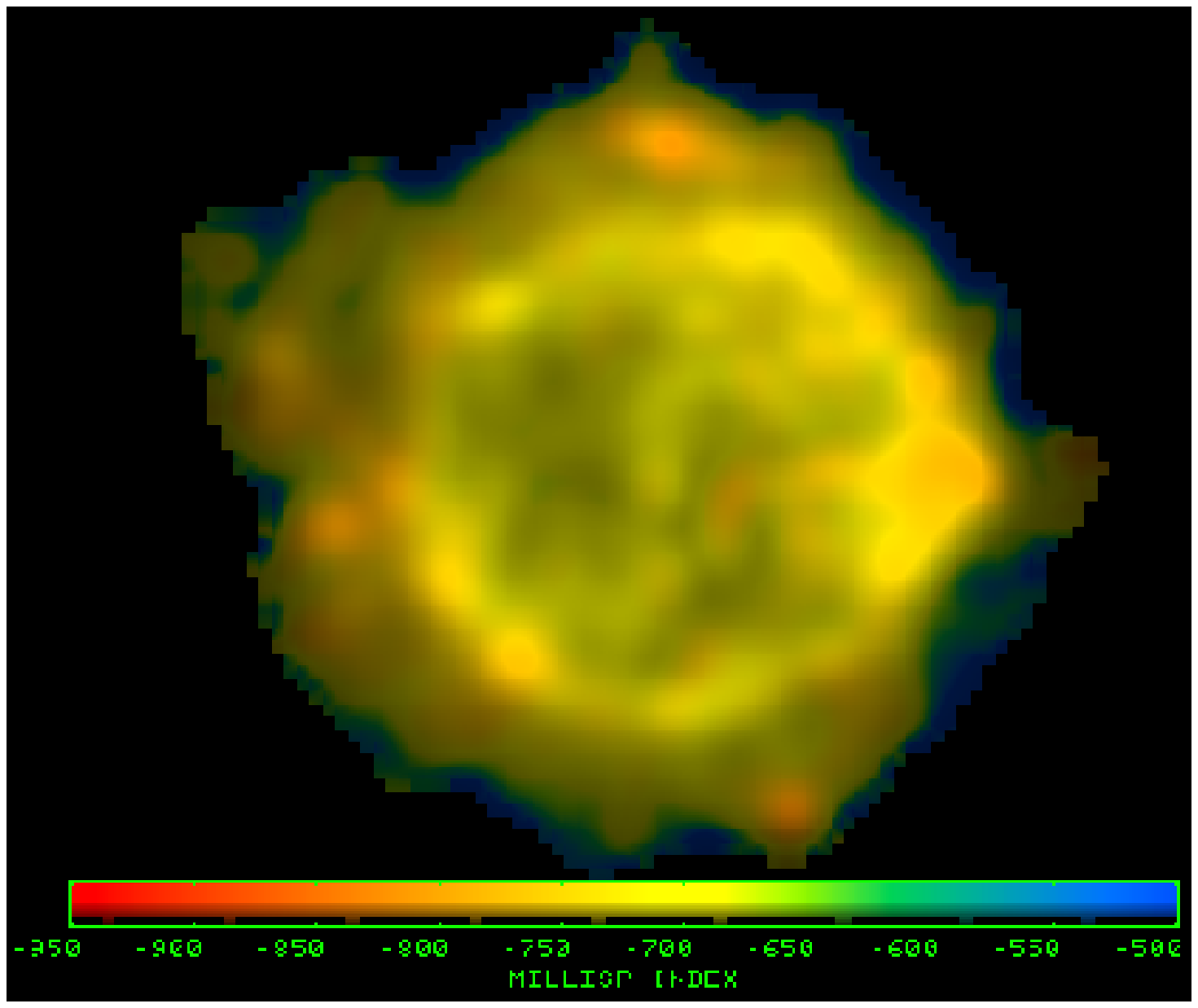}
\includegraphics[height=0.41\textwidth]{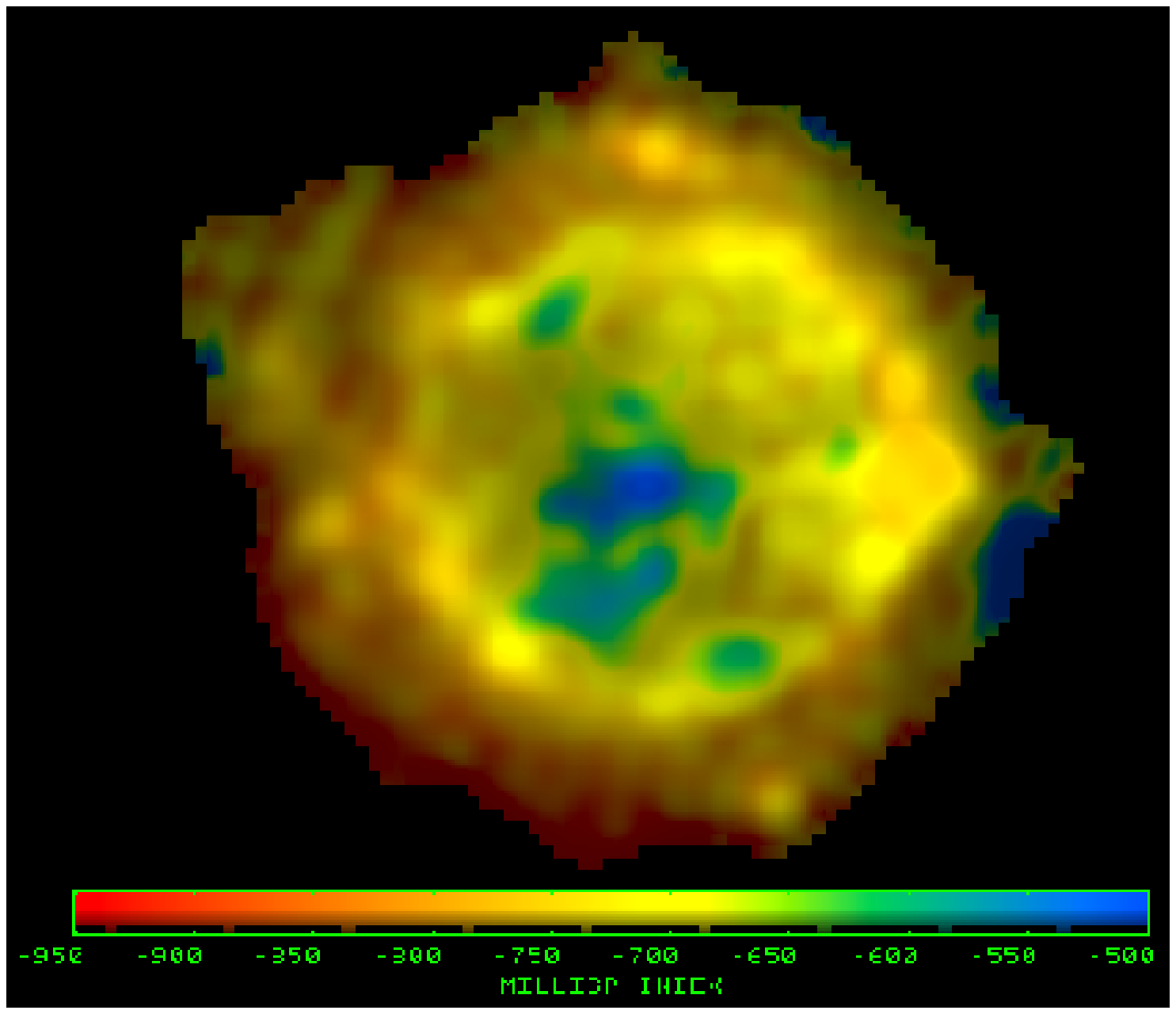}
\caption{Spectral index images between 330~MHz and 1.4~GHz (top) and between 
74~MHz and 330~MHz (bottom) made from the 1997-1998 data set.  Only spectral 
index data from regions greater than 10-$\sigma$ on the 330~MHz image are 
shown.  The 330~MHz image is used as a brightness channel to illuminate the 
spectral index colors.  The resolution is 18\farcs5.  
\label{spixims}}
\end{figure}

The 1997-1998 74~MHz, 330~MHz, and 1.4~GHz images at 18\farcs5 resolution were 
combined to form the spectral index images presented in Figure~\ref{spixims}.  
Only spectral index data from regions greater than 10-$\sigma$ on the 330~MHz 
image were retained.  These images show similar patterns in spectral index as 
those presented by \citet{kpd95} at 25\arcsec~resolution but at higher 
signal-to-noise for the images tied to 74~MHz because of the full complement 
of antennas available.  As with the higher resolution spectral index image in 
Figure~\ref{ptspix}, the central features in the $\alpha_{74}^{330}$ image are 
notably flatter than the Bright Ring.  One distinct advantage at lower spatial 
resolution is that the signal-to-noise is high enough to probe the spectral 
index of the diffuse emission as well as the clumped emission.  The range of 
spectral indices between 330~MHz and 1.4~GHz (Figure~\ref{spixims} left) is 
$\alpha_{330}^{1.4}\approx$-0.85 to -0.65 and is typical of that observed at 
higher frequencies \citep{arl91}.  The range of spectral indices between 
74~MHz and 330~MHz (Figure~\ref{spixims} right) is 
$\alpha_{74}^{330}\approx$-0.95 to -0.5, which is slightly larger than that 
between 330~MHz and 1.4~GHz.  Both images in Figure~\ref{spixims} are plotted 
using the same spectral index scale in order to better demonstrate the 
spectral flattening effect in the central regions of the $\alpha_{74}^{330}$ 
image.  At 18\farcs5 resolution, the central region is not as flat as at 
9\arcsec~resolution.  This could be because the lower resolution smooths out 
large variations and also because of the lower dynamic range of the 
9\arcsec~resolution image.

As noted by \citet{kpd95}, the spectral indices for features on the Bright 
Ring are about the same from high frequencies to low frequencies.  At the 
center of Cas~A, however, there is a significant flattening of the spectral 
index between 330~MHz and 74~MHz.  Given that the spectral flattening is 
confined to the center of the 74~MHz image, there is likely an absorbing 
medium in or near Cas~A that affects the observed low frequency emission but 
not the observed high frequency emission.

\subsection{Spectral Index Difference ($\Delta\alpha$)}

The spectral index variations observed at high frequencies reflect the 
different energy spectra of accelerated electron populations.  These different 
energy spectra may be due to the acceleration at the forward and reverse 
shocks and to intrinsic breaks in the underlying electron energy spectrum 
\citep{arl91}.  The spectral index variations observed at low frequencies are 
due to both the accelerated electron populations and effects from free-free 
absorption.  

\begin{figure}[t!]
\centering
\epsscale{1.17}
\plotone{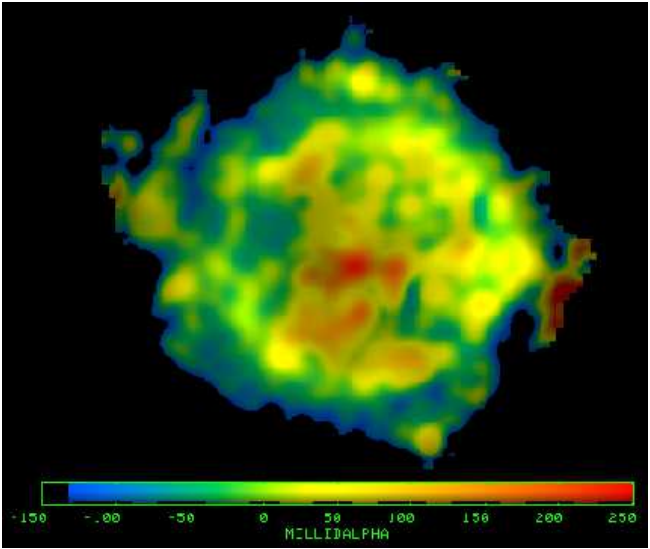}
\caption{Spectral index difference ($\Delta\alpha$ or dalpha) between 
$\alpha_{74}^{330}$ and $\alpha_{330}^{1.4}$. Only spectral index data from 
regions greater than 10-$\sigma$ on the 330~MHz image are shown.
The 330~MHz image is used as a brightness channel to illuminate the 
$\Delta\alpha$ colors.  The resolution is 18\farcs5.  The orange and red 
colors indicate locations with free-free absorption.
\label{dalpha}}
\end{figure}

In order to isolate spectral changes due only to free-free absorption, we 
assume that there is no curvature in the \emph{emitted} synchrotron spectrum 
at low frequencies.  Thus, when we subtract the two spectral index images, the 
resulting image, $\Delta\alpha=\alpha_{74}^{330}-\alpha_{330}^{1.4}$, shows only 
variations due to free-free absorption.  Our $\Delta\alpha$ image is shown in 
Figure~\ref{dalpha} with the same signal-to-noise cutoff as the spectral index 
images.  The 330~MHz image was again used to illuminate the colors.  Typical 
$\Delta\alpha$ values for regions with good S/N ratio vary from -0.1 to 0.25 
where positive $\Delta\alpha$ indicates absorption and statistical variations 
due to noise are of order $\pm$0.01 to 0.02.  Most of the free-free absorption 
is confined to the middle of Cas~A with typical $\Delta\alpha$ values of 
0.1-0.25.  This is the same absorption morphology noted by \citet{kpd95} who 
found a peak $\Delta\alpha$=0.3 with their 25\arcsec~resolution images.  

\section{Comparison with Infrared Emission from Unshocked Ejecta}
\label{sec:clumps}

\begin{figure*}[t!]
\centering
\epsscale{0.5}
\includegraphics[width=0.48\textwidth]{fig8a.ps}
\includegraphics[width=0.45\textwidth]{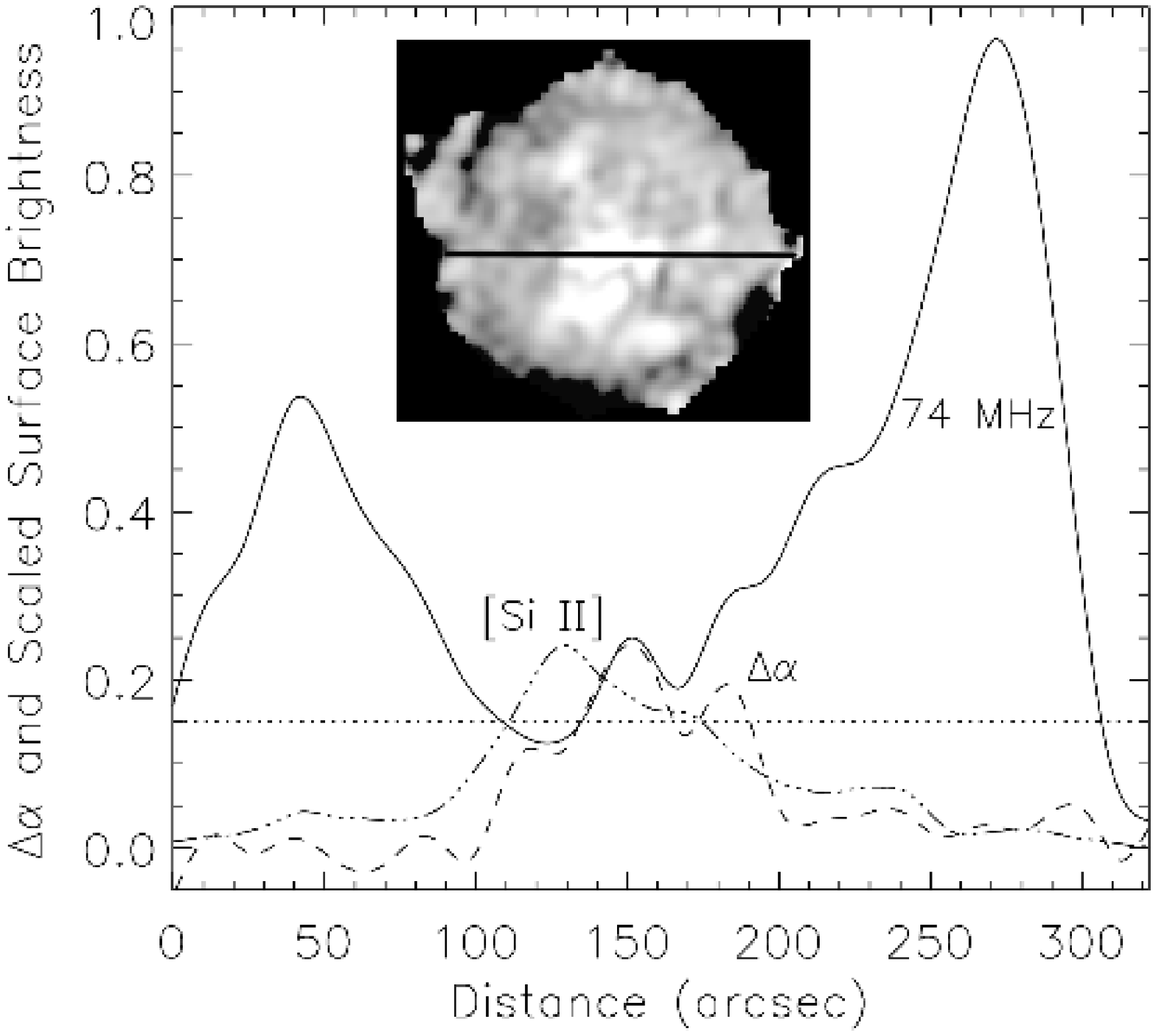}\\
\includegraphics[width=0.48\textwidth]{fig8c.ps}
\includegraphics[width=0.47\textwidth]{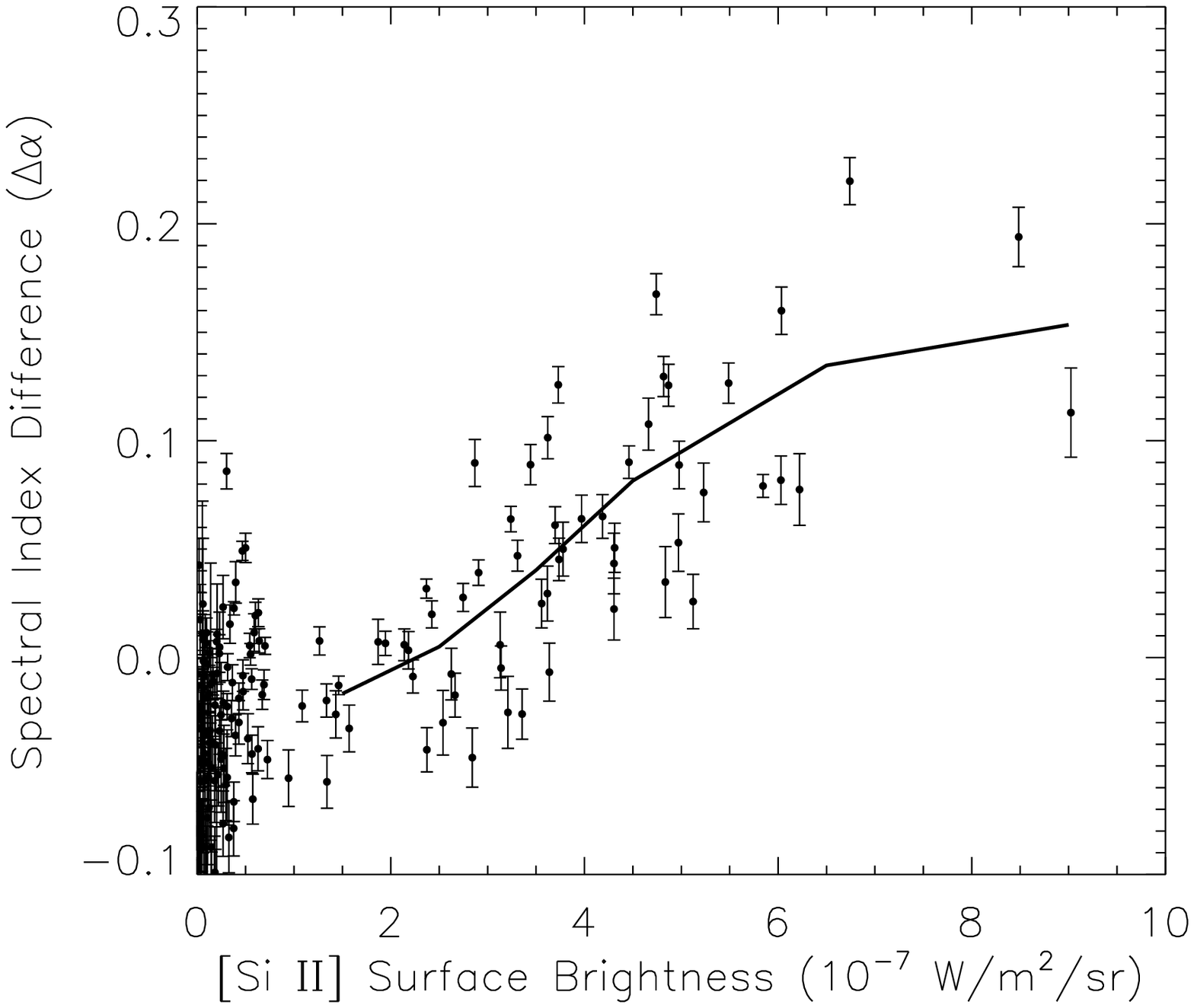}
\caption{Top left: $\Delta\alpha$ contours plotted over infrared [\ion{Si}{2}] 
emission in greyscale at a spatial resolution of 18\farcs5.  
Positive $\Delta\alpha$ values (solid contours) indicate free-free 
absorption.  The contour levels correspond to $\Delta\alpha$ values of: 
-0.05, 0.00, 0.05, 0.10, 0.15, 0.20, 0.25, and 0.30 and the signal-to-noise 
cutoff for the contours is 10-$\sigma$ on the 330~MHz image. Top right: Slice 
profiles taken at the location indicated on the inset of the $\Delta\alpha$ 
image.  The $y$-axis indicates the value of $\Delta\alpha$ and the 74~MHz and 
[\ion{Si}{2}] profiles have been scaled to show correlations between features 
in the images.  The dotted line indicates the ``average'' $\Delta\alpha$ 
chosen for our analysis.  Bottom left: Same as top left except with blanking 
applied to exclude shocked ejecta.  Bottom right: $\Delta\alpha$ vs. 
[\ion{Si}{2}] surface brightness from only unshocked ejecta.  The solid line 
indicates the running mean of the $\Delta\alpha$ values and shows a general 
increase of absorption with [\ion{Si}{2}] surface brightness.
\label{siabs}}
\end{figure*}

One of the most exciting discoveries with \emph{Spitzer} 
was the presence of emission from unshocked ejecta in Cas~A.  This emission is 
most prominent in the lines of [\ion{O}{4}] and [\ion{Si}{2}] and less 
prominent for [\ion{S}{3}] and [\ion{S}{4}] \citep{err06}.  The density and 
temperature conditions derived for this material based on the infrared line 
ratios confirms that it is indeed cold, low density, unshocked, photoionized 
ejecta \citep{srd09}.  In order to confirm the hypothesis of \citet{kpd95} 
that the low frequency free-free absorption seen in Cas~A is due to unshocked 
ejecta, we compare the $\Delta\alpha$ image to the infrared [\ion{Si}{2}] 
image.  As shown in the top left panel of Figure~\ref{siabs}, strong free-free 
absorption ($\Delta\alpha\ga0.1$) corresponds well to bright [\ion{Si}{2}] 
emission ($\ga5\times10^{-7}$ W m$^{-2}$ sr$^{-1}$).  However, there are bright 
[\ion{Si}{2}] regions that have low or no associated absorption. 

There is not a one-to-one correlation between free-free absorption and 
[\ion{Si}{2}] emission for three primary reasons.  First, while 
the [\ion{Si}{2}] image is a good representative of the emission observed 
from the unshocked ejecta in the \emph{Spitzer} data, it also contains 
emission from shocked ejecta.  As the ejecta at large radius cross the reverse 
shock, become heated, and contribute to the Bright Ring emission, the 
free-free absorption drops drastically due to the temperature dependence.  We 
can filter the [\ion{Si}{2}] image using the infrared [\ion{Ar}{2}] image to 
blank those regions corresponding to shocked ejecta.  The remaining 
[\ion{Si}{2}] emission, shown in the bottom left panel of Figure~\ref{siabs}, 
is dominated by unshocked ejecta and, when plotted against $\Delta\alpha$ 
(bottom right panel of Figure~\ref{siabs}), shows a good correlation to the 
low frequency absorption.  The second reason for scatter in the correlation 
between the free-free absorption and [\ion{Si}{2}] emission is that the 
[\ion{Si}{2}] image provides no information about other unshocked components 
that may be present but not detected, such as higher ionization states of Si- 
and O-rich ejecta or the presence of Fe-rich ejecta, for example.  Finally, 
the $\Delta\alpha$ image is a function of both the column density of the 
absorbing medium and the distribution of radio emitting regions along the line 
of sight.  The presence of clumped radio emission biases the $\Delta\alpha$ 
calculation to higher or lower values depending on whether the clump is on the 
far or near side of the unshocked ejecta in Cas~A.  The greater the brightness 
disparity between the clump and the diffuse background, the stronger the 
bias.  We show this case in the top right panel of Figure~\ref{siabs} where we 
plot slice profiles for $\Delta\alpha$ and 74~MHz and [\ion{Si}{2}] surface 
brightnesses.  The peak $\Delta\alpha$ occurs at the location of a 
particularly bright emission clump, but has no obvious correlation to any 
distinct [\ion{Si}{2}] features.  

We therefore conclude that the free-free absorption is clearly correlated to 
the observed emission from the unshocked ejecta and is a tracer of 
this material as presumed by \citet{kpd95}.  Since free-free absorption 
depends on the temperature and density of the absorbing medium, our low 
frequency radio data provide a means of determining the conditions in the 
unshocked ejecta.  When coupled with assumptions about geometry, we can also 
calculate the total mass of the observed unshocked ejecta.  We carry out these 
calculations in the following sections.

\section{Density and Mass of the Unshocked Ejecta}

\subsection{Free-Free Optical Depth}
\label{optdep}

Our use of thermal absorption to probe the properties of the unshocked ejecta 
in Cas~A relies on simplifying assumptions. Our measurement of $\Delta\alpha$, 
discussed above, can yield an equivalent free-free optical depth, but relies 
on our knowledge of the line-of-sight brightness distribution of the 
nonthermal, illuminating radiation from Cas~A's shell.  The high spatial 
resolution images in Figure~\ref{2asec} show that Cas~A consists of both 
clumpy and diffuse components.  We assume the smooth emission has radial 
symmetry, consistent with shell decomposition models \citep{fwp80,gkr01}, so 
that equal contributions come from the front and back side of the shell along 
any line of sight.  As discussed in \S\ref{sec:clumps}, the presence of clumps 
will bias our measurement of $\Delta\alpha$.  However, it is almost impossible 
to avoid clumped emission in Cas~A, as Figure~\ref{2asec} shows.  We could try 
using a higher frequency image to account for the clumps, but without a 
priori information about which clumps are in front of or behind the absorbing 
medium, we would introduce more uncertainties into the analysis.  Therefore, 
we choose to select a representative absorption figure that effectively 
averages out the contamination from clumps.  In the top right panel of 
Figure~\ref{siabs}, we show the absorption profile taken along the slice 
indicated in the inset.  The peaks and valleys of the central absorption 
($\Delta\alpha$) are correlated with 74~MHz surface brightness variations, as 
expected.  Typical $\Delta\alpha$ values in the center of Cas A range from 
0.1-0.25 so we choose by eye $\Delta\alpha$=0.15 as a compromise between high 
and low absorption towards the center of Cas~A. Since this is a somewhat 
arbitrary choice, we will investigate in \S\ref{sec:vary} how the final mass 
and density estimates for the unshocked ejecta depend on the value of 
$\Delta\alpha$.  

We also assume that the cold ejecta lie only in the interior of Cas~A, and 
therefore only affect the 74~MHz emission from the back side of the shell, 
again consistent with three-dimensional models \citep{drs10}. We further assume 
the 330~MHz emission is completely unabsorbed -- consistent with our spectral 
index measurements from 330~MHz to 5~GHz. 

Based on the above picture, we derive in Appendix~A the free-free optical depth 
$\tau$ as a function of $\Delta\alpha$:
\begin{equation}
\tau = -\ln[2 (10^{a \Delta \alpha} -0.5)]
\end{equation}
where $a=\log(74/330)$.  In this formulation, spectral index is 
defined as $\alpha_{74}^{330}=\frac{\log(S_{74}/S_{330})}{\log(74/330)}$ and 
therefore is negative for a synchrotron spectrum.  Using $\Delta\alpha$=0.15 
results in a $\tau$ of about 0.51.

\subsection{Emission Measure of the Unshocked Ejecta}

The free-free absorption $\tau$ can be characterized by the following formula 
\citep[chap.~9; see also Appendix B]{rw00}:

\begin{eqnarray} 
&\tau_{\nu}=&3.014\times10^{4} \left(\frac{Z^2}{f}\right)\left(\frac{\nu}{\mathrm{MHz}}\right)^{-2}\left(\frac{T_e}{\mathrm{K}}\right)^{-3/2}\left(\frac{EM}{\mathrm{pc\,cm}^{-6}}\right) \nonumber \\
\label{taueq1}  
& & \times \ln \left[49.55 Z^{-1}\left(\frac{T_e}{\mathrm{K}}\right)^{3/2} \left(\frac{\nu}{\mathrm{MHz}}\right)^{-1}\right] 
\end{eqnarray}
where $Z$ is the average atomic number of the ions dominating the cold 
ejecta, $f$ is the electron to ion ratio, $\nu$ is the radio frequency in MHz 
at which the absorption is observed, $T_e$ is the temperature in K of the 
ionized absorbing medium, and $EM$ is the emission measure in units of 
pc cm$^{-6}$.  Emission measure is defined as:
\begin{equation}
\label{emden} 
EM \equiv \int n_e^2 dl 
\end{equation}
where $n_e$ is the electron density of the gas in units of cm$^{-3}$ and $dl$ 
is the path length along the line-of-sight measured in parsecs.  Inverting 
Equation~\ref{taueq1} to solve for $EM$ yields:
\begin{equation}
EM=\frac{3.318\times10^{-5} \left(\frac{\tau_{\nu} f}{Z^2}\right)\left(\frac{\nu}{\mathrm{MHz}}\right)^2 \left(\frac{T_e}{\mathrm{K}}\right)^{3/2}}{\ln \left[49.55\,Z^{-1}\left(\frac{T_e}{\mathrm{K}}\right)^{3/2} \left(\frac{\nu}{\mathrm{MHz}}\right)^{-1}\right]}.
\end{equation}

\subsection{Composition of the Absorbing Medium}
\label{sec:composition}

To interpret $EM$, we first need to place realistic constraints on both $Z$ and 
$f$. Based on our observed correlation between $\Delta\alpha$ and 
infrared [\ion{Si}{2}] emission, and the emission from unshocked 
ejecta in the infrared lines of [\ion{O}{4}], [\ion{S}{3}], and [\ion{S}{4}] 
observed with \emph{Spitzer} \citep{err06,e09}, we assume the absorption is 
dominated by Si- and O-rich ejecta, including Si, S, Ar, Ca, Mg, Ne, and O. To 
determine the relative abundances of these species in Cas~A, we rely on X-ray 
derived abundances indicating that the ejecta in Cas~A are O-dominated 
\citep{wbv02} and, based on stellar composition models, are typically more 
than a million times more abundant than hydrogen \citep{ww86}. Based on the 
abundances quoted in \citet{wbv02} from \emph{XMM-Newton} observations of 
Cas~A we adopt a weighted average $Z$ of 8.34.

\begin{figure}
\centering
\epsscale{1.25}
\plotone{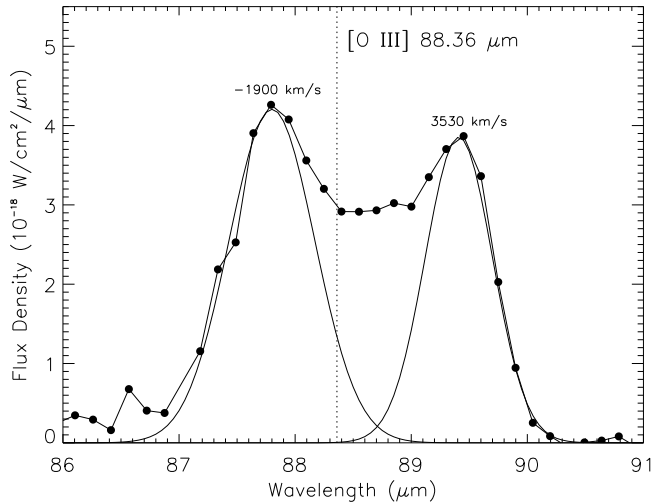}
\caption{ISO spectrum of the center of Cas~A taken with the LWS instrument 
in 1996.  The dotted line marks the rest wavelength of [\ion{O}{3}].  Using 
the rest wavelength of 88.36~$\mu$m for [\ion{O}{3}] results in Doppler 
velocities of the fitted Gaussian components as indicated.  These velocities 
correspond to the known velocity structure of the unshocked ejecta 
\citep{drs10}.   We note that [\ion{Fe}{2}], with a rest wavelength of 
87.38~$\mu$m, also falls into this same wavelength range, but the data are 
completely consistent with the [\ion{O}{3}] origin given the known velocity 
structure. 
\label{isospec}}
\end{figure}

Spectra from \emph{Spitzer} and \emph{ISO} \citep[see Figure \ref{isospec},][]{upc97,ds10} indicate that both [\ion{O}{3}] and [\ion{O}{4}] are observed from 
the unshocked ejecta.  While models of unshocked ejecta in Cas~A predict 
higher ionization states to be present, the material responsible for the 
observed infrared emission and most of the thermal absorption is in these 
lower ionization states \citep{e09}.  Therefore, we adopt an electron/ion 
ratio, $f$, of 2.5.

At the observing frequency of 73.8~MHz, the last parameter to consider 
is $T_e$.  \citet{kpd95} assumed, rather arbitrarily, $T_e$=1000~K in their 
calculation.  A much better estimate of the temperature in the unshocked 
ejecta comes from the analysis of the infrared emission lines.  In order to 
produce strong [\ion{O}{4}] and [\ion{Si}{2}] lines but weak or absent 
[\ion{S}{4}] and [\ion{Ar}{3}] lines in the \emph{Spitzer} data, $T_e$ must 
be $\sim$100-500~K \citep{e09}.  Therefore, we will adopt $T_e$=300~K.

Parameterized in terms of our values for $\tau_{\nu}$, $f$, $Z$, $T_e$, and 
$\nu$, the emission measure becomes: 
\begin{equation}
\label{emparam}
EM=\frac{17.21\left(\frac{\tau_{\nu}}{0.51}\right)\left(\frac{f}{2.5}\right)\left(\frac{Z}{8.34}\right)^{-2}\left(\frac{T_e}{300\,\mathrm{K}}\right)^{3/2}\left(\frac{\nu}{73.8\,\mathrm{MHz}}\right)^2}{\ln\left[418.3\left(\frac{T_e}{300\,\mathrm{K}}\right)^{3/2}\left(\frac{Z}{8.34}\right)^{-1}\left(\frac{\nu}{73.8\,\mathrm{MHz}}\right)^{-1}\right]}.
\end{equation} 

\subsection{Geometry of Cas~A}
\label{sec:geometry}

In order to estimate density from the emission measure and then estimate the 
total mass in the unshocked ejecta, we must assume a geometry for the 
absorbing medium.  \citet{drs10} and \citet{ird10} showed that the unshocked 
ejecta, as traced by [\ion{O}{4}] and [\ion{Si}{2}] emission, are concentrated 
onto two thick sheets interior to the reverse shock -- one front and 
one rear.  The two sheets are separated by a much lower density region, where 
only a little emission is seen.  One might expect unshocked C-rich and Fe-rich 
ejecta at various locations interior to the reverse shock as well based on 
stratification models of the ejecta \citep{err06}, however these species are 
not observed from the unshocked ejecta either in the \emph{Spitzer} IRS 
spectra or the ISO LWS spectra, which could indicate that they are of lower 
density than the O-rich or Si-rich material, or simply absent \citep{e09}.

\begin{figure}
\centering
\epsscale{1.17}
\plotone{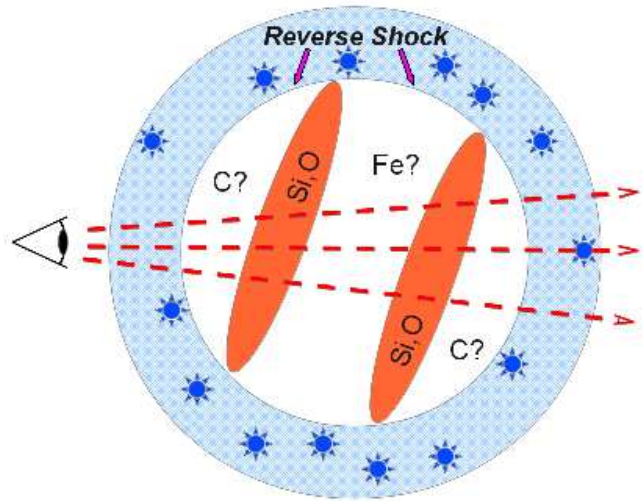}
\caption{Simplified drawing of the geometry of the shocked and unshocked
material in Cas~A. The shaded, circumferential ring represents the region 
bounded on its interior and exterior by the reverse shock and the blast wave, 
respectively. It consists of smooth (blue shaded) and clumped (blue stars)  
synchrotron emitting components.  The two orange ellipsoids represent 
thermally absorbing, unshocked ejecta emitting infrared lines of Si and O. C 
and Fe, depicted as lying exterior and interior to the Si and O emitting 
material, respectively, have not been detected but might be present at lower 
densities. \label{cartoon}}
\end{figure}

A simplified drawing of the geometry involved is shown in Figure~\ref{cartoon}.
As explained in \S\ref{optdep}, the emission from shocked material in 
Cas~A can be modeled as a spherically symmetric shell consisting of 
smooth and clumped components.  The clumpy shocked material may be ejecta, 
circumstellar material, or filaments associated with the forward or reverse 
shock.  The unshocked ejecta are confined to two thick sheets in the 
interior.  The high spectral resolution \emph{Spitzer} mapping shows that the 
sheets are composed of thin filaments (1\farcs5 or 0.03 pc thick) with several 
filaments along any given line-of-sight \citep{ird10}.  Therefore, we estimate 
the combined thickness of the front and back sheets to be of order 10$\arcsec$ 
which translates to $L=$0.16~pc at the distance to Cas~A of 3.4~kpc 
\citep{drs10,rhf95}.  The sheets intersect the reverse shock at a radius of 
about 90$\arcsec$ (or 1.48~pc), so that the total volume of absorbing material 
is $V=\pi R^2 L =$1.1 pc$^3$.  

The geometry can be modified further by the use of a clumping factor.  
We know from the optical images of shocked ejecta in Cas~A that there is a 
great deal of structure on small scales with typical knot sizes between 
0\farcs2 and 0\farcs4 \citep{fmc01}.  It is not known how much of this 
clumping was present prior to reverse shock passage.  The effect of clumping 
is to reduce the total volume of material while at the same time increasing 
the density of the material.  In order to determine the effect of clumping on 
our calculations, we simply modify each of our radius or thickness measures by 
the clumping factor ($CL$ or $CR$) so that when $C=1$ there is no clumping and 
when $C<1$ there is some degree of clumping.  Our default value of $C$ is to 
assume no further clumping beyond the filamentary structures identified by 
\citet{ird10}.   

\subsection{Electron Density and Mass}

If we assume that the density is constant, then emission measure defined in 
Equation~\ref{emden} simplifies to $EM=n_e^2CL$ where $L$ is the combined 
thickness of the unshocked ejecta sheets.  In Appendix~C we show electron 
density fully parameterized in terms of $\tau_{\nu}$, $f$, $Z$, $T_e$, $\nu$, 
$C$, and $L$, however the variables in the denominator are suppressed by both 
a natural logarithm and a square root. Therefore we show here a simplified 
version of the parameterized formula in order to show how electron density 
depends on these various quantities:
\begin{eqnarray}
\label{neapprox}
n_e \approx & 4.23 \left(\frac{\tau_{\nu}}{0.51}\right)^{1/2}\left(\frac{f}{2.5}\right)^{1/2}\left(\frac{Z}{8.34}\right)^{-1}\left(\frac{T_e}{300\,\mathrm{K}}\right)^{3/4} \nonumber \\
& \left(\frac{\nu}{73.8\,\mathrm{MHz}}\right)C^{-1/2}\left(\frac{L}{0.16\,\mathrm{pc}}\right)^{-1/2}.
\end{eqnarray}
Thus, our estimate of electron density in the unshocked ejecta is 4.2~cm$^{-3}$.

The total mass of unshocked ejecta is simply $M=\rho V$ where $\rho$ is the 
mass density of the ions, $\rho=Zm_pn_e/f$, and $m_p$ is proton mass.  If we 
substitute Equation~\ref{nefull} for $n_e$, assume the two thick sheet 
geometry, and parameterize mass in terms of $\tau_{\nu}$, $f$, $Z$, $T_e$, 
$\nu$, $C$, $R$, and $L$ then, as shown in Appendix~C, the dependence on $Z$ 
becomes minuscule as the only surviving $Z$-term is mitigated by the natural 
logarithm and square root in the denominator.  As we did for electron density, 
we show here a simplified parameterized equation for the total mass of the 
unshocked ejecta expressed in solar masses:
\begin{eqnarray}
M \approx & 0.39 \left(\frac{\tau_{\nu}}{0.51}\right)^{1/2}\left(\frac{f}{2.5}\right)^{-1/2}\left(\frac{T_e}{300\,\mathrm{K}}\right)^{3/4} \nonumber \\
& \left(\frac{\nu}{73.8\,\mathrm{MHz}}\right)C^{5/2}\left(\frac{R}{1.48\,\mathrm{pc}}\right)^2\left(\frac{L}{0.16\,\mathrm{pc}}\right)^{1/2}.
\end{eqnarray}
We therefore estimate a total mass in unshocked ejecta of 0.39~M$_{\sun}$  
which is significantly improved compared to the 19~M$_{\sun}$ estimated by 
\citet{kpd95}.

\subsection{Effect of Varying Parameters}
\label{sec:vary}

\begin{figure}
\centering
\epsscale{1.17}
\plotone{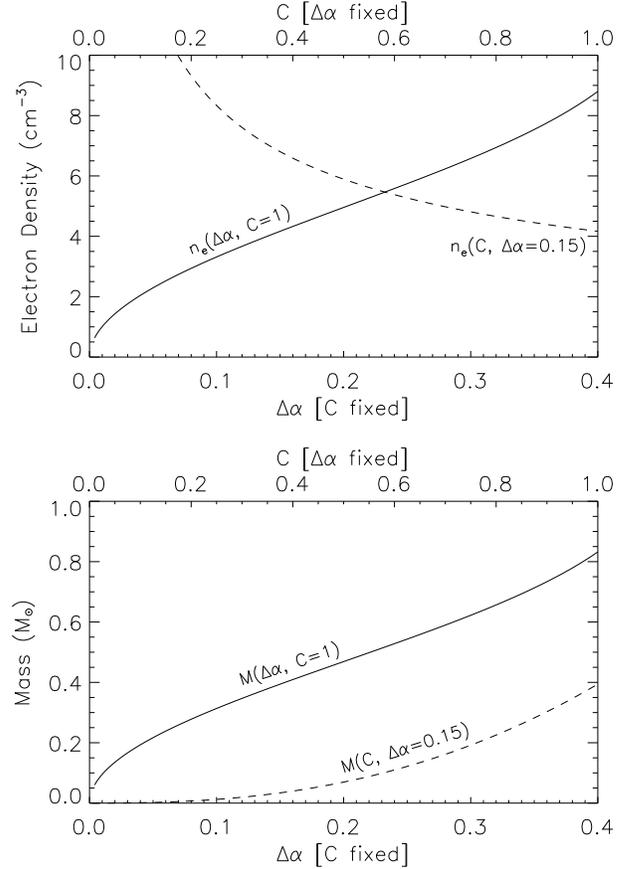}
\vspace{-6mm}
\caption{Plots of Equation~\ref{nefull} (top) and Equation~\ref{mfull} 
(bottom) as functions of $\Delta\alpha$ (via $\tau$) and clumping factor.  
For each curve, one of the variables are held constant as indicated.  Each of 
the $x$-axes is independent with $\Delta\alpha$ indicated on the bottom 
$x$-axis and clumping ($C$) indicated on the top $x$-axis in each plot. 
\label{vardac}}
\end{figure}

We now delve into the effect our choices of $\Delta\alpha$, $C$, $T_e$, $f$, 
$Z$, and geometry have on the final electron density and mass estimates.  We 
start with our measurement of $\Delta\alpha$ since the values are constrained 
by the radio observations.  In Figure~\ref{vardac} we plot 
Equations~\ref{nefull} and \ref{mfull} as functions of $\Delta\alpha$ 
(via $\tau$). If $\Delta\alpha$ is varied from 0.1 to 0.3, typical of the 
$\Delta\alpha$ values found near the middle of Cas~A, the electron 
density and mass estimates change by a factor of two (3.3 to 6.6 cm$^{-3}$ and 
0.31 to 0.62 M$_{\sun}$, respectively).  Even if ``very strong'' absorption 
($\Delta\alpha$=0.4) were observed uniformly across the center of Cas~A, the 
maximum mass estimate would still be below 1~M$_{\sun}$.

Also shown in Figure~\ref{vardac} are Equations~\ref{nefull} and \ref{mfull} 
plotted as functions of clumping.  Recall that $C=1$ indicates no clumping, 
while smaller values of $C$ increase the amount of clumping in the ejecta.  As 
clumping increases, the electron density increases and the total mass estimate 
decreases.  For all modest clumping estimates, the electron density remains 
below 10 cm$^{-3}$ and the total mass remains below 1~M$_{\sun}$.

\begin{figure}
\centering
\epsscale{1.17}
\plotone{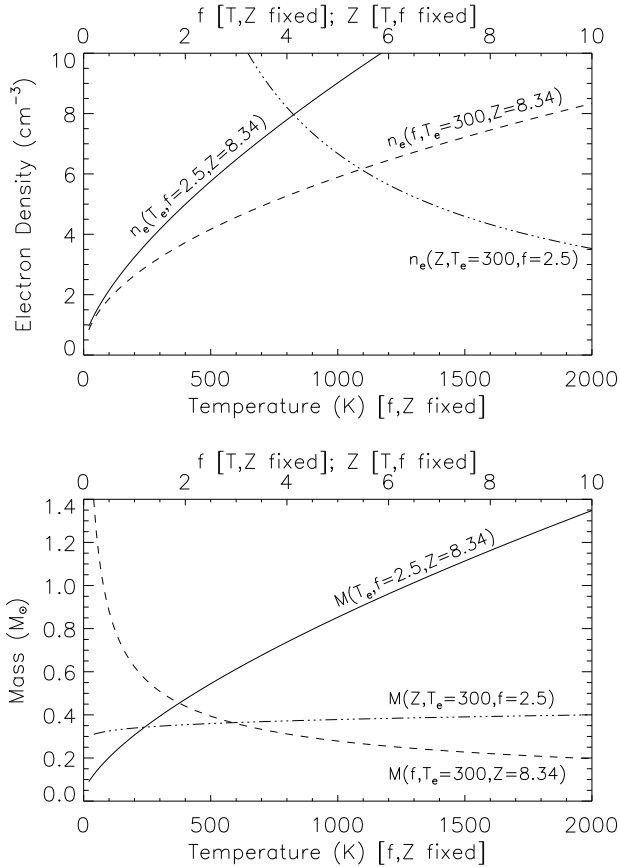}
\vspace{-6mm}
\caption{Plots of Equation~\ref{nefull} (top) and Equation~\ref{mfull} 
(bottom) as functions of $T_e$, $f$, and $Z$.  For each curve, two of the 
variables are held constant as indicated.  Each of the $x$-axes is 
independent with temperature indicated on the bottom $x$-axis and atomic 
number ($Z$) and electron/ion ratio ($f$) indicated on the top $x$-axis in 
each plot. \label{variables}}
\end{figure}

The values of $T_e$, $f$, and $Z$ depend on emission models and observations 
that may be affected by components that are missing because we have no probe 
for them, such as higher ionization states of Si \citep{e09}.  To illustrate 
the effect of varying these quantities, we plot Equations~\ref{nefull} and 
\ref{mfull} in Figure~\ref{variables}.  For each of the plotted curves, two of 
the values are held constant as indicated.  All reasonable estimates of 
$T_e$, $f$, and $Z$ require that the electron density be less than 
10~cm$^{-3}$.  While the total unshocked ejecta mass is much more strongly 
influenced by temperature and electron/ion ratio than atomic number, 
plausible variations in all three variables constrain the mass below 
1~M$_{\sun}$.

The final effect we would like to address is that of the geometry.  While the 
two thick sheet geometry for the unshocked ejecta is supported by the 
morphology of the infrared [\ion{Si}{2}] and [\ion{O}{4}] emission, and the 
correlation between the $\Delta\alpha$ map and [\ion{Si}{2}] is quite good, 
this is not a one-to-one correlation.  Other species possibly in different 
locations along the line-of-sight might be responsible for some of the 
observed thermal absorption.  Therefore, we consider here a spherical geometry 
in which the unshocked ejecta are modeled as a filled sphere of absorbing 
material interior to the bright radio-emitting ring.  Since the emission 
measure derives from the $\Delta\alpha$ measurement, it does not change, 
however the path length through the absorbing medium does.  If we use 
$L=2.96$~pc through the center of Cas~A so that $R=L/2=1.48$~pc as before, and 
we assume that $n_e$ is constant, then the electron density from 
Equations~\ref{neapprox} and \ref{nefull} becomes 0.98 cm$^{-3}$ and the total 
mass of the unshocked ejecta becomes 1.12~M$_{\sun}$ as derived in 
Equation~\ref{msph}.  A clumping factor would increase the electron density 
and decrease the total mass estimate.

\section{Discussion}

\subsection{Improvement over Previous Mass Estimate}
\label{sec:oldwork}

As mentioned previously, \citet{kpd95} estimated a total mass of 19~M$_{\sun}$ 
for the unshocked ejecta.  The dominant reason for this high mass value is 
that the free-free optical depth equation used by \citet{kpd95} was for a 
hydrogenic gas, rather than an oxygenic gas, as appropriate for a core-collapse 
SNR.  In addition, \citet{kpd95} made a number of assumptions about other 
model parameters that significantly affected their final mass values.  
Specifically, their temperature was 1000~K compared to our value of 300~K.  
They used their maximum $\Delta\alpha$ of 0.3 compared to our more modest 
estimate of 0.15.  They correctly assumed the ejecta were dominated by oxygen, 
but incorrectly assumed that the oxygen was singly ionized.  \citet{kpd95} 
also assumed a spherical ejecta distribution rather than our two-thick-sheet 
morphology.  Thus, the confluence of parameter values chosen, and using an 
incorrect formula for free-free optical depth, resulted in a very high mass 
estimate in the earlier work.

\subsection{Comparison with Theoretical Expectations for Density and Mass}
\label{sec:theory}

The density and temperature conditions in the unshocked ejecta and the total 
mass of this material play a key role in many aspects of supernova remnant 
physics.  \citet{lh03} and \citet{hl03,hl12} have incorporated 
results from their analysis of the million-second observation of Cas~A with 
\emph{Chandra} into hydrodynamic simulations of supernova remnant evolution 
that account for factors such as expansion into a stellar wind and 
acceleration of the reverse shock.  Their models, designed to match the 
dynamics of Cas~A, result in a range of total and shocked ejecta masses, 
which are then partitioned into the various elemental species based on their 
derived abundances from the X-ray data.  Their most recent prediction for the 
total mass of ejecta that are currently unshocked is 0.18-0.3~M$_{\sun}$, 
depending on the model chosen.  Which species dominate the composition of the 
unshocked ejecta is debatable since the explosion is known to have been 
asymmetric \citep{rfs11}.  \citet{hl12} argue that there is little Fe left in 
the unshocked ejecta based on several factors including the lack of infrared 
Fe emission from the unshocked ejecta \citep{srd09}, the location of the 
X-ray-emitting Fe-rich ejecta at the same or larger radius than the Si-rich 
ejecta \citep{hrb00} and the carbon atmosphere of the neutron star 
\citep{hh09}.  Our estimate of 0.39~M$_{\sun}$ for the unshocked ejecta is 
near the upper limit predicted by \citet{hl12}, however reasonable variations 
of our parameters, such as increasing the clumping factor, can lead to lower 
values of total unshocked ejecta mass in line with the prediction.  

We can also compare the density of the unshocked ejecta to the hydrodynamic 
model of \citet{hl12}.  Their ejecta model assumes a uniform density core 
surrounded by a power-law envelope.  The density of the core as a function of 
time is given by:
\begin{equation}
\label{meanrho}
\rho_{\mathrm{core}}=\left(\frac{M_{ej}}{v_{\mathrm{core}}^3}\right)\left(\frac{3}{4 \pi}\right)\left(\frac{n-3}{n}\right)t^{-3}
\end{equation}
where $M_{ej}$ is the total mass of the ejecta and the factor $(n-3)/n$ gives 
the fraction of the ejecta mass in the core \citep{lh03,hl12,tm99}.  
$v_{\mathrm{core}}$ is the velocity of the boundary between the core and the 
power-law envelope.  For an $n=10$ power-law envelope and $M_{ej}=$3~M$_{\sun}$, 
\citet{hl12} report $v_{\mathrm{core}}\approx9000$ km s$^{-1}$.  At an age of 330 
years, $\rho_{\mathrm{core}}\approx1\times10^{-24}$ g cm$^{-3}$.  Our ``filled 
sphere'' geometry provides a convenient comparison because it, too, assumes a 
uniform density ejecta distribution throughout the interior of the SNR with no 
clumping.  For $n_e=0.98$ cm$^{-3}$, $f$=2.5, and $Z$=8.34, we derive 
$\rho_{\mathrm{sph}}=5.5\times10^{-24}$ g cm$^{-3}$, which is only a 
factor of 5 larger than \citet{hl12}. 

The ejecta in Cas~A are distributed in a complex fashion and the unshocked 
ejecta show signs of filamentary behavior in the high spectral resolution 
\emph{Spitzer} data \citep{ird10}.  For any given emission measure, if the 
material is tied up in clumps, the final density estimate will be higher and 
the final mass estimate would be lower than when the material is distributed 
uniformly along the line-of-sight.  In order to match our mass estimate with 
that of \citet{hl12}, this would argue for the unshocked ejecta being 
primarily distributed in the two-shell geometry.  Any unshocked C-rich or 
Fe-rich ejecta must then be of such low density that it makes little 
contribution to the observed thermal absorption -- or one or both of these 
components are missing entirely.  

Our calculations favor a low density and low temperature environment in the 
unshocked ejecta consistent with the analysis of the \emph{Spitzer} infrared 
data \citep{e09}.  Temperatures $\lesssim$500~K are not unexpected given the 
rapid expansion of the SNR, which could conceivably cool the ejecta to 
$\sim$10s of Kelvin, and the fact that the infrared flux densities infer 
unshocked dust temperatures of about 35~K \citep[see e.g.][]{nkt10}.  
Expectations for the density of the unshocked ejecta are not clear.  While 
hydrodynamic models can predict an average density, as in 
Equation~\ref{meanrho}, the clumping of the ejecta renders the average 
somewhat meaningless.  If the unshocked ejecta we detect 
via thermal absorption and mid-infrared emission eventually radiate strongly 
in the X-rays upon shock heating, then our density estimate is in line with 
that expectation.  The electron densities observed in the X-ray-emitting 
(shocked) ejecta are $\sim$10s~cm$^{-3}$ which would be consistent with the 
standard strong shock compression ratio of 4.  On the other hand, much more 
dense (1000s~cm$^{-3}$) shocked ejecta are observed optically 
\citep[see e.g.][]{ck79} indicating that the unshocked ejecta must have a much 
denser, clumped component in order to reproduce the observed range of shocked 
ejecta densities \citep{mfc04}.  A high degree of clumping might arise due to 
the instabilities associated with the explosion reverse shock which forms 
inside of the star in the first few hours after the explosion begins 
\citep{hw94,jwh09}.  Similarly, turbulence and strong clumping are associated 
with expanding Fe-Ni bubbles in the inner ejecta \citep{lms93,bbr01}.  The 
expansion of Fe-Ni bubbles naturally renders the innermost Fe ejecta of very 
low density, consistent with the lack of detected infrared Fe emission in the 
unshocked ejecta.  

\subsection{Comparison to SN1006}
\label{sec:1006}

The only other confirmed detection of unshocked ejecta in an SNR is that of 
SN1006 where high-velocity absorption features were observed in ultraviolet 
lines of \ion{Fe}{2} and \ion{Si}{2} \citep{wls83,hfw97}.  High resolution 
spectra showed no detectable unshocked \ion{Si}{3} or \ion{Si}{4} and thus the 
bulk of the unshocked Si is accounted for with \ion{Si}{2} absorption
\citep{wlh05,hfb07}.  The total Si mass calculated from the derived column 
densities (0.25 M$_{\sun}$) is roughly comparable with the Si mass predicted 
from some white dwarf carbon deflagration models (0.16 M$_{\sun}$) and the 
derived Si mass from X-ray data (0.2 M$_{\sun}$) \citep{hfw97}.  

Accounting for the unshocked Fe in SN1006 has proven to be more difficult.  
\citet{hfw97} argue that the unshocked Fe absorption should be dominated 
by \ion{Fe}{2} because the photoionization cross section is five times that of 
\ion{Si}{2}.  However, the total mass in \ion{Fe}{2} (0.029 M$_{\sun}$) is less 
than 1/10 of that expected from carbon deflagration models (0.3 M$_{\sun}$) 
\citep{hf88,hfw97}.  The absence of \ion{Fe}{2} absorption along some 
lines-of-sight indicate that there is likely no large-scale mixing or 
overturning of Fe ejecta, so the bulk of the Fe must still be interior to the 
reverse shock \citep{wlh05}.  There is the possibility that the Fe is tied up 
in higher ionization states \citep{hf88}, but then the lack of \ion{Si}{3} and 
\ion{Si}{4} absorption from the unshocked ejecta is puzzling \citep{hfw97}.  
Therefore, it seems that some of the unshocked ejecta in SN1006 are 
still unaccounted for.

In Cas~A, there is a similar ``problem'' between the strong emission from 
[\ion{Si}{2}] and the lack of any detectable infrared Fe emission from the 
unshocked ejecta in the \emph{Spitzer} IRS bands or the ISO LWS spectra.  At 
the density we infer for the unshocked ejecta, [\ion{Si}{2}] should be much 
stronger than [\ion{Fe}{2}] and [\ion{Fe}{3}], however [\ion{Fe}{5}] should be 
comparable to the [\ion{Si}{2}] \citep{e09}, which is not observed 
\citep{srd09}.  As explained in \S\ref{sec:theory}, the Fe discrepancy in 
Cas~A may be resolved if the bulk of the Fe ejecta have already passed through 
the reverse shock and the remaining interior Fe is of lower density than the 
unshocked Si- and O-rich ejecta.  However, we need a better knowledge of 
Cas~A's progenitor and the expected Fe yields for a Type IIb explosion of such 
a star in order to determine if all of the Fe ejecta in Cas~A are truly 
accounted for.

One other comparison we can make to SN1006 is of the density in the unshocked 
ejecta.  \citet{hfw97} estimate 
$\rho_{\mathrm{FeII}}\sim4\times10^{-27}$~g~cm$^{-3}$ at the center of their 
unshocked ejecta profile with an estimate of about half that figure for the 
current unshocked mass density of \ion{Si}{2}.  If we scale the mass density 
of the (O-rich) unshocked ejecta in Cas~A by 27 to account for the difference 
in age of the two SNRs and thus the dilution of the density due to expansion, 
then $\rho_{\mathrm{sph}}^{1000}=2\times10^{-25}$ g cm$^{-3}$.  Therefore the 
\emph{observed} unshocked ejecta in Cas~A are about 100 times more dense than 
in SN1006.  Given that Cas~A is the result of a Type IIb explosion and SN1006 
resulted from a Type Ia explosion, stark differences in ejecta density are to 
be expected due to differing stellar compositions, explosion mechanisms and 
resulting nucleosynthesis, initial ejecta density profiles, instabilities in 
the ejecta that may lead to clumping, and circumstellar environments.  The 
question is are the unshocked ejecta in Type Ia SNRs systematically less dense 
than in core-collapse remnants?  The extension of low frequency radio analysis 
to other SNRs could be used to test this hypothesis.  Furthermore, since the 
absorption is frequency dependent, working at frequencies lower than 74~MHz 
will allow the detection of lower density ejecta in Cas~A and other SNRs.

\subsection{Implications for Lower Frequency Spectrum}

\citet{hk09} revisited the low frequency, integrated spectrum of Cas~A 
utilizing legacy VLA and Long Wavelength Development Array 74 MHz
measurements. We consider here how the thermal absorption discussed in this 
paper relates to the turnover in the integrated spectrum of Cas~A at much 
lower frequencies, as originally documented by \citet{bgp77}. Given the 
current limited angular resolution of observations towards Cas~A below 74~MHz, 
our discussion is necessarily limited.

Examination of the \citet{bgp77} Figure~1a indicates the integrated spectrum of
Cas~A begins to turn over below 38~MHz. The \citet{bgp77} modeled spectrum
(epoch 1965) which did not fit for a turnover, predicts $\sim$93,000~Jy at
10~MHz. The lowest reliable flux density measurement at that frequency is the
Penticton 10~MHz flux density of 28,000~Jy (epoch 1965.9, i.e. 
contemporaneous, \citealt{b67}), which is a factor of 3.3 below the expected 
value.  (This corresponds to a 10~MHz free-free optical depth by a uniform 
external absorber of $\sim$1.2, an oversimplification since some of the
absorption must be internal.) As described in Section~\ref{optdep}, an average
free-free optical depth for the unshocked ejecta at 74~MHz is $\sim$0.5, which
rapidly inflates to $\gg$1 at 10~MHz, i.e. completely opaque. If the absorbing
ejecta blocked $\sim$30\% of the back surface of Cas~A's nonthermal emission,
$\sim$85\% of Cas~A's 10~MHz emission, or $\sim$79,000~Jy, should escape the
remnant. We can account for the remaining deficit by an intervening thermal
absorber with $\tau_{10\,\mathrm{MHz}} \sim -\ln (28,000/79,000) \sim 1$.
Thus, a simplistic scenario has $\sim$79,000~Jy of Cas~A's 10~MHz emission 
emerging from the remnant after being attenuated by $\sim$15\% from the 
unshocked ejecta. Thereafter the emerging emission is further attenuated by 
an intervening ionized component of the ISM with 10~MHz optical depth $\sim$1.

The distribution of low density gas in the interstellar medium was constrained
by \citet{k89}, using the observed, patchy thermal absorption inferred from 
the low frequency, integrated spectra of Galactic SNRs. (Several of those 
systems have since been resolved with legacy VLA data \citep{llk01,blk05}.) The 
measured optical depths at 30.9~MHz typically ranged from 0.1-1, though some 
remnants showed no turnovers while others were more heavily absorbed. The 
nature of the absorption was consistent with higher frequency recombination 
line observations that postulated extended \ion{H}{2} region envelopes (EHEs)
associated with normal \ion{H}{2} regions \citep{ana86}. The inferred ISM 
10~MHz free-free optical depth unity discussed above, and required to further 
attenuate Cas~A's emerging emission, scales to $\sim$0.1 at 30.9 MHz.  
Thus the low frequency turnover in the integrated spectrum of Cas~A is 
qualitatively consistent with a combination of intrinsic absorption from 
unshocked ejecta and extrinsic absorption by EHEs located along the line of 
sight. In retrospect the \citet{bgp77} low frequency turnover would be 
conspicuous in its absence based on what we now know about ionized gas in 
both the ISM and Cas~A.

\section{Future Work}
\label{sec:future}

The radio detection of thermally absorbing, unshocked ejecta inside a young
Galactic SNR by the legacy VLA 74~MHz system only scratches the surface of
possibilities for emerging low frequency systems with greater capabilities.
With the transition to the fully digital electronics and new correlator of the 
VLA, the narrowband 74 and 330~MHz legacy systems have been replaced with an 
improved, broad-band ``Low Band'' receiving system \citep{ckh11}. The new, 
single receiver system brings significant improvements over its two narrow 
band predecessors, and will become available to the VLA user community in 
2014.

For broad band systems probing below 74~MHz, e.g. LOFAR Low Band 
\citep{lofar}, or LWA, the frequency dependence of thermal absorption 
significantly enhances the effect. Gradually tuning to lower frequencies 
as the absorption grows can break the optical depth degeneracy between 
temperature and emission measure, and deconvolve the radial superposition of 
nonthermal emitting and thermal absorbing regions. Lower frequency 
observations with sufficient angular resolution (e.g. LOFAR Low Band) can test 
our qualitative assessment of the relative contributions of intrinsic and 
extrinsic thermal absorption contributing to the observed turnover in Cas~A's 
integrated spectrum below 38~MHz. It is also important to constrain the 
emission at frequencies above 74~MHz just prior to the onset of absorption, as 
offered e.g. by LOFAR High Band, MWA, the GMRT, and VLA Low Band.

\citet{hk09} also confirmed shorter scale (5-10 yr) temporal variations at 
74~MHz consistent with previous reports in the literature for low frequencies 
\citep{ep75, vr05}. These have generally been attributed to the interaction of 
the forward shock with an inhomogeneous CSM or ISM. However some variation 
must also arise from inside Cas~A, as the reverse shock advances into the 
unshocked ejecta and affects its opacity. Together with the secular decrease 
due to adiabatic expansion \citep{s60}, the overall temporal behavior is 
necessarily complex, and requires higher resolution, broad band low frequency 
measurements to understand. We suggest that Cas~A's low frequency flux density 
should be monitored routinely across a range of lower frequencies; for 
instruments such as LOFAR or LWA this would require only a few minutes of 
observations every few weeks or months and would allow a much better 
determination of the range of timescales over which such variations occur.

\section{Conclusions}

We have imaged Cas~A from 5~GHz to 74~MHz in all four configurations of 
the Legacy VLA with follow-up observations at 74 and 330~MHz with the 
legacy VLA+PT link.  Our spatially resolved spectral index maps confirm the 
interior spectral flattening measured earlier, but at higher signal-to-noise 
and resolution. Comparison with \emph{Spitzer} infrared spectra confirms the 
earlier hypothesis that the spectral flattening is due to thermal absorption 
by cool, unshocked ejecta photoionized by X-ray radiation from Cas~A's 
reverse shock. We use the spectral flattening to measure the free-free optical 
depth.

Next, using priors of electron temperature, atomic number, and electron to 
ion ratios, we derive an emission measure from the measured optical depth. 
With an assumed geometry, informed from three-dimensional modeling based on 
higher frequency studies, we use the emission measure to place constraints on 
both the density and total mass of the unshocked ejecta.

We consider modest, physically plausible variations in both our priors and the 
assumed geometry, and find that the effect on the total mass is relatively 
modest, varying by a factor of about two. Furthermore, our derived total mass 
is consistent with recent model predictions \citep{hl12}. After accounting for 
the relative ages of Cas~A and SN1006, our derived mass density is much higher 
than found in SN1006, not unexpected since Cas~A (Type IIb) and SN1006 
(Type Ia) emerged from two fundamentally different supernova explosion types.  
However, if there is a systematic difference in unshocked ejecta density for 
core collapse vs. Type Ia SNRs, low frequency radio data can be used to test 
this hypothesis.

Finally, we consider the contribution of the intrinsic thermal absorption to 
the known turnover of Cas~A's integrated spectrum at much lower frequencies. 
We find that the intrinsic thermal absorption from the unshocked ejecta, 
combined with extrinsic absorption from a known, patchy distribution of low 
density ISM gas, are completely consistent with the low frequency turnover.

The promise of the emerging instruments is expanding the population of SNRs, 
young and old, that can be probed for intrinsic and extrinsic thermal 
absorption and shock acceleration variations beyond pathologically bright 
sources like Cas~A. More generally, the seemingly ubiquitous detection of 
resolved thermal absorption by the 74~MHz legacy VLA against the Galactic 
background \citep{nhr06} and towards, discrete non thermal sources 
\citep[e.g. see][]{llk01,blk05,cdb07} confirms the phenomena will continue to 
emerge as a powerful tool for low frequency astrophysics.

\acknowledgements
The VLA is operated by the National Radio Astronomy Observatory, which is a 
facility of the National Science Foundation, operated under cooperative 
agreement by Associated Universities, Inc.  All sub-GHz systems on the VLA 
have been developed cooperatively between the National Radio Astronomy 
Observatory and the Naval Research Laboratory.  Basic research in radio 
astronomy at the Naval Research Laboratory is supported by 6.1 base funding.  
Partial funding for this research at West Virginia Wesleyan College was 
provided by Chandra Grant GO0-11089X and by the NASA-West Virginia Space Grant 
Consortium.

\emph{Facility:} \facility{VLA}

\appendix
\section{Derivation of $\tau-\Delta\alpha$ Relation}

The definition of spectral index  between 74 and 330~MHz is 
\begin{equation}
\alpha_{74}^{330}=\frac{\log(S_{74}/S_{330})}{\log(74/330)},
\end{equation}
where $\alpha$ is negative for nonthermal emission. We also define 
$a=\log\left(\frac{74}{330}\right)$.

Assuming no curvature in the emitted synchrotron spectrum, we expect ($e$) 
the spatially resolved spectral index $\alpha$ to remain unchanged from above 
330~MHz to between 330 and 74~MHz, defining $\alpha_e$. $\alpha_o$ is then 
defined as the observed ($o$) spectral index due to absorption.

The difference between observed and expected spectral index is then: 
$\Delta \alpha=\alpha_o-\alpha_e$ ($\Delta \alpha$ is positive), so that:
\begin{equation}
\Delta \alpha = \frac{\log(S_{o74}/S_{330}) - \log(S_{e74}/S_{330})}{a} = \frac{\log(S_{o74}/S_{e74})}{a}
\end{equation}

To obtain observed and expected 74~MHz flux density, we assume equal 
synchrotron emission from the front and back sides ($S_f=S_b$) of Cas~A:
\begin{eqnarray}
S_e=S_f + S_b = 2S_f\\
S_o=S_f+S_f e^{-\tau} = (1+e^{-\tau})S_f\\
\frac{S_o}{S_e}=\frac{1+e^{-\tau}}{2}
\end{eqnarray}

Substituting for $\frac{S_o}{S_e}$ we have:
\begin{eqnarray}
a \Delta \alpha = \log\left(\frac{S_{o74}}{S_{e74}}\right)=\log\left(\frac{1+e^{-\tau}}{2}\right)\\
2 (10^{a \Delta \alpha}) = 1 + e^{-\tau}\\
\tau = -\ln[2 (10^{a \Delta \alpha} -0.5)] 
\label{taueqn}
\end{eqnarray}

We note that Equation~\ref{taueqn} is slightly different than that published 
in \citet{kpd95} due to a typographical error.

\section{Calculation of Free-Free Optical Depth $\tau$}

In this section, we follow the derivation in chapter 9 of \citet{rw00}.
 
The definition of free-free optical depth ($\tau$) is:
\begin{equation}
\tau_{\nu}=-\int \kappa_{\nu} dl
\end{equation} 
where
\begin{equation}
\kappa_{\nu} = \frac{4 Z^2 e^6}{3 c} \frac{n_i n_e}{\nu^2} \frac{1}{\sqrt{2 \pi (m k T_e)^3}} \left<g_{ff}\right>
\end{equation}
is the absorption coefficient and the Gaunt factor is
\begin{equation}
\left<g_{ff}\right> = \ln\left[\left(\frac{2 k T_e}{\gamma m_e}\right)^{3/2} \frac{m_e}{\pi \gamma Z e^2 \nu}\right].
\end{equation}
$Ze$ is the ion charge, $n_i$ and $n_e$ are number density of ions and 
electrons, respectively.  $c$ is the speed of light, $\nu$ is the observation 
frequency, $m_e$ is the electron mass, $k$ is Boltzmann's constant, 
$\gamma=1.781$, and $T_e$ is the effective temperature of the gas which must 
be $>20$~K.

For classic \ion{H}{2} regions one normally assumes a hydrogenic gas where 
$Z=1$ and $n_e=n_i$, but that is not the case here \citep{ww86}.  Let 
$f=n_e/n_i$ and use emission measure $=EM \equiv \int n_e^2\,dl$ and express 
$T_e$ in K, $\nu$ in MHz, and $EM$ in pc cm$^{-6}$ so that:
\begin{equation} \tau_{\nu}=3.014\times10^{4} \left(\frac{Z^2}{f}\right)\left(\frac{\nu}{\mathrm{MHz}}\right)^{-2}\left(\frac{T_e}{\mathrm{K}}\right)^{-3/2}\left(\frac{EM}{\mathrm{pc\,cm}^{-6}}\right)\ln \left[49.55 Z^{-1}\left(\frac{T_e}{\mathrm{K}}\right)^{3/2} \left(\frac{\nu}{\mathrm{MHz}}\right)^{-1}\right]. 
\end{equation}

\section{Full Parameterized Equations for Electron Density and Total Mass}

The electron density follows from Equation~\ref{emden} and our preferred 
geometry as $n_e=\sqrt{EM/CL}$.  Substituting Equation~\ref{emparam} into this 
expression yields:
\begin{equation}
\label{nefull}
n_e = \frac{10.4 \left(\frac{\tau_{\nu}}{0.51}\right)^{1/2}\left(\frac{f}{2.5}\right)^{1/2}\left(\frac{Z}{8.34}\right)^{-1}\left(\frac{T_e}{300\,\mathrm{K}}\right)^{3/4}\left(\frac{\nu}{73.8\,\mathrm{MHz}}\right)C^{-1/2}\left(\frac{L}{0.16\,\mathrm{pc}}\right)^{-1/2}}{\left\{\ln\left[418.3\left(\frac{T_e}{300\,\mathrm{K}}\right)^{3/2}\left(\frac{Z}{8.34}\right)^{-1}\left(\frac{\nu}{73.8\,\mathrm{MHz}}\right)^{-1}\right]\right\}^{1/2}}
\end{equation}
where $n_e$ is in units of cm$^{-3}$.

The total mass in unshocked ejecta is calculated from 
$M=(Zm_pn_e/f)(\pi C^3 R^2 L)$, where $m_p$ is proton mass.  Substituting 
Equation~\ref{nefull} for $n_e$ and absorbing constant terms results in the 
following equation for total mass in solar masses:
\begin{equation}
\label{mfull}
M = \frac{0.97 \left(\frac{\tau_{\nu}}{0.51}\right)^{1/2}\left(\frac{f}{2.5}\right)^{-1/2}\left(\frac{T_e}{300\,\mathrm{K}}\right)^{3/4}\left(\frac{\nu}{73.8\,\mathrm{MHz}}\right)C^{5/2}\left(\frac{R}{1.48\,\mathrm{pc}}\right)^2\left(\frac{L}{0.16\,\mathrm{pc}}\right)^{1/2}}{\left\{\ln\left[418.3\left(\frac{T_e}{300\,\mathrm{K}}\right)^{3/2}\left(\frac{Z}{8.34}\right)^{-1}\left(\frac{\nu}{73.8\,\mathrm{MHz}}\right)^{-1}\right]\right\}^{1/2}}.
\end{equation}

If a spherical geometry is assumed, then 
$M=(Zm_pn_e/f)(\frac{4}{3}\pi C^3 R^3)$ which becomes:
\begin{equation}
\label{msph}
M = \frac{2.73 \left(\frac{\tau_{\nu}}{0.51}\right)^{1/2}\left(\frac{f}{2.5}\right)^{-1/2}\left(\frac{T_e}{300\,\mathrm{K}}\right)^{3/4}\left(\frac{\nu}{73.8\,\mathrm{MHz}}\right)C^{5/2}\left(\frac{R}{1.48\,\mathrm{pc}}\right)^3\left(\frac{L}{2.96\,\mathrm{pc}}\right)^{-1/2}}{\left\{\ln\left[418.3\left(\frac{T_e}{300\,\mathrm{K}}\right)^{3/2}\left(\frac{Z}{8.34}\right)^{-1}\left(\frac{\nu}{73.8\,\mathrm{MHz}}\right)^{-1}\right]\right\}^{1/2}}.
\end{equation}

\end{document}